\documentclass[12pt,hidelinks]{article}

\usepackage{amssymb,amsmath,amsthm,amsfonts,changepage,eurosym,geometry,ulem,graphicx,caption,color,setspace,comment,footmisc,caption,natbib,pdflscape,subcaption,array,hyperref,soul,bm,float,booktabs,longtable}
\usepackage[labelsep=period]{caption} 
\usepackage[section]{placeins}
 
\theoremstyle{definition}
\newtheorem*{defn*}{\protect\definitionname}
\providecommand{\definitionname}{Definition}

\theoremstyle{plain}

\newcolumntype{L}[1]{>{\raggedright\let\newline\\arraybackslash\hspace{0pt}}m{#1}}
\newcolumntype{C}[1]{>{\centering\let\newline\\arraybackslash\hspace{0pt}}m{#1}}
\newcolumntype{R}[1]{>{\raggedleft\let\newline\\arraybackslash\hspace{0pt}}m{#1}}

\begin{document}

\begin{titlepage} 
\title{Dissertation Paths: Advisors and Students in the Economics Research Production Function\thanks{We thank our exceptionally committed and hard-working research assistants, Alison Fang, Merilin Martinson, Isabel Munoz, and Tianyuan (Margaret) Zheng for two years of painstaking work. Thanks also go to Glenn Ellison, Drew Fudenberg, and Tobias Salz for comments, to RFBerlin for research funding, to Reyn van Ewijk for letting us use his server, and to Liz Braunstein for invaluable help with digital EconLit. This essay is dedicated to the memory of Oberlin College Professor Hirschel Kasper, a pioneer in the study of economics education and undergraduate advisor to Josh Angrist. Kasper's report on economics education was published as \cite{kasper1991}.}}
\author{Joshua Angrist\thanks{MIT and NBER. Email: \href{mailto:angrist@mit.edu}{angrist@mit.edu}} \quad Marc Diederichs\thanks{ROCKWOOL Foundation Berlin and University of Passau. Email: \href{mailto:marc.diederichs@uni-passau.de}{marc.diederichs@uni-passau.de}} }
\date{December 2024}
\maketitle
\begin{abstract}
\singlespacing
\noindent 
Elite economics PhD programs aim to train graduate students for a lifetime of academic research. This paper asks how advising affects graduate students’ post-PhD research productivity. Advising is highly concentrated: at the eight highly-selective schools in our study, a minority of advisors do most of the advising work. We quantify advisor attributes such as an advisor’s own research output and aspects of the advising relationship like coauthoring and research field affinity that might contribute to student research success. Students advised by research-active, prolific advisors tend to publish more, while coauthoring has no effect. Student-advisor research affinity also predicts student success. But a school-level aggregate production function provides much weaker evidence of causal effects, suggesting that successful advisors attract students likely to succeed–without necessarily boosting their students’ chances of success. Evidence for causal effects is strongest for a measure of advisors’ own research output. Aggregate student research output appears to scale linearly with graduate student enrollment, with no evidence of negative class-size effects. An analysis of gender differences in research output shows male and female graduate students to be equally productive in the first few years post-PhD, but female productivity peaks sooner than male productivity.

\par 
\end{abstract} 
\onehalfspacing
\setcounter{page}{0}
\thispagestyle{empty}
\end{titlepage}
\pagebreak 
\newpage

\onehalfspacing

\graphicspath{ {C:\Users\madieder\Dropbox\PhD Research Productivity\code\Marc Analyses\figures} }

\section{Introduction}

The most selective doctoral programs in economics promise to teach their students to write and publish journal articles reporting on their research. Although many economics PhD students land in non-research positions in consulting, finance, or government, elite program curriculum is research-oriented. This can be seen in the programs' lengthy reading lists, demanding research exercises such as second- and third-year papers, and the time students are expected to invest in their job market papers and thesis chapters. Stratospherically--and increasingly--selective economics PhD programs target bright and ambitious students who appear committed to and well-prepared for a career of academic economics research.\footnote{The MIT Department of Economics, long a flagship for economics graduate education, admitted 38 of 794 PhD applicants in 2014 and just 27 of 895 PhD applicants in 2024. MIT Economics' current 3\% acceptance rate is below that of even the most selective Ivy League colleges.}

These scholarly aspirations notwithstanding, half of elite economics PhDs, graduates of Harvard, MIT, Stanford and the like, publish next to nothing in the 6 years following degree completion, while only 5-10\% publish more than a paper or two \citep{conley2014research}.\footnote{These statistics are for 1986-2000 PhD graduates of 30 top departments.} Elite schools employ stellar faculty with lengthy, influential research careers. Surprisingly, the \cite{conley2014research} data suggest that graduates of good-but-not elite programs (classified according to widely-used rankings), like CMU, Rochester and San Diego, publish about as well as do graduates from Harvard, MIT, and Stanford. Why do so few highly-selected elite program graduates follow the path to research success taken by their extraordinarily successful advisors? What aspects of economics advisee training might be changed or enhanced so as to boost graduate student success and total research output?  

These questions motivate our study of the economics PhD education production function at elite universities. The principal production inputs in this function are the faculty who teach and advise graduate students, along with aspects of the advising relationship that faculty and students develop together. We aim to measure features of the advising relationship and to link these features with students’ research success. Graduate education has many features; we focus here on those most obviously tied to research. Specifically, we look at advisors' own research success and aspects of an advisor's advising history such as the number of past advisees and the scholarship of former students.  We also consider measures of research affinity such as whether an advisee's doctoral thesis cites advisor work and the extent of coauthoring between advisors and their students. Notably, we ignore teaching and graduate classes.  This reflects our view that students in the elite programs we study are exposed to broadly similar levels of coursework. Advising relationships, by contrast, are highly idiosyncratic even within programs.

Our analysis starts with an ambitious data collection effort linking doctoral dissertations with advisor characteristics and measures of recent graduates' research productivity for cohorts of PhD students completing an economics degree since 1994. We focus on eight elite schools, some of which have multiple programs training economics research PhDs. These schools are, in alphabetical order, Berkeley, Chicago, Harvard, MIT, Northwestern, Princeton, Stanford, and Yale. Our linked sample contains roughly 8,000 graduates from economics departments and economics-adjacent programs like those in business and public policy schools granting economics-related PhDs.  By the early 2000s, the graduates in our sample account for roughly 20\% of research articles published in the 137 or so most cited journals indexed by the AEA's bibliographic database Econlit (their advisors likewise account for around 20\%). The students of interest authored roughly half of articles published in top-6 economics journals in 2020.\footnote{Journal lists come from \cite{angrist2020inside}, which identifies the journals most cited by the \emph{American Economic Review}. Top-6 journals include the usual top 5 plus the \emph{Review of Economics and Statistics}. For a sample graduating 1987-92, \cite{collins2000publishing} likewise report a preponderance of top program graduates' research publications in 36 highly-cited and top-5 economics journals.} 

Our investigation of the economics graduate education production function is related to earlier analyses of economics graduate education and to studies of research training in other fields.  \cite{waldinger2010quality}, for instance, shows that Nazi Germany's expulsion of Jewish and politically unreliable mathematics faculty degraded PhD student success in affected departments. \cite{corsini2022makes} examines the effect of advisor characteristics on STEM PhD student success, focusing on advisor gender as a causal factor. \cite{gaule2018advisor} and \cite{neumark1998women} investigate the importance of gender matching between advisors and graduate students in the sciences and economics, respectively. Following a broader analysis of student research output, we look briefly at gender effects and interactions in the PhD student research production function as well. Our gender analysis is related to \citet{card2020referees,iaria2024gender}, and \citet{sarsons2021gender}, among others.

Also related, \cite{hilmer2009fishes,hilmer2011you} attempt to distinguish the effects of advisor scholarship from program prestige effects in a sample of economics PhDs. \cite{garcia2020predicting} does something similar using recently-available data from the \href{http://repec.org/}{\textcolor{blue}{Repec}} economics research paper repository. \cite{athey2007does}  examine effects of graduate school performance indicators like grades on job placement. Building on this work, our analysis examines the role of advisor networks and considers estimates with and without control for school fixed effects.  Natural sciences PhD students appear to benefit from advising and mentoring by highly visible and productive superstar scholars \citep{li2019early}. At the same time, an analysis of life sciences students suggests PhD graduates are more successful when their dissertation research synthesizes work from areas an advisor or mentor's field \citep{lienard2018intellectual}.  We likewise aim to assess the importance of research affinity for economics PhDs.  


Our analysis is distinguished from related earlier work by our large recent sample and by the scope of the explanatory variables considered as inputs in the economics PhD research production function.\footnote{\cite{buchmueller1999graduate} reports regression estimates of the effect of pre-graduation publication on later research output computed using a non-representative survey sample of two older cohorts.} Importantly, we also tackle the problem of selection bias and systematic sorting in estimates of advisor effects on student success. Successful advisors, however defined, likely attract successful students. If so, the relationship between an advisor's past students' success and a new advisee's research productivity need not be causal. A school-level analysis mitigates this by asking what happens to average success at say, Princeton and Berkeley, when these schools employ more or fewer prolific advisors.  The resulting estimates show surprisingly little evidence of superstar advisor effects at the school level.  A similar analysis considers the effect of program size, asking whether larger and therefore less selective programs face declining returns to scale in research success.  Results here suggest that the best way to increase a program's total research impact, defined as the number of high-quality publications its graduates produce, is to increase the number of PhD students the program graduates.  

The next section sketches aspects of our data set construction, with details covered in an accompanying data appendix.  Section \ref{sec:researchoutput} presents a descriptive overview of the research output generated by the PhD students in our sample. This section also looks briefly at gender gaps and gender matching in the advising relationship. Section \ref{sec:advising} discusses estimates of the relationship between advisor attributes--advisor research prominence, past advisee numbers, and former student success--and current advisee publication success. This section also includes an examination of the role played by advisor-advisee coauthoring and research topic affinity in determining student publication rates. Section \ref{sec:aggregate} looks at aggregate research productivity by school in a grouped data instrumental variables (IV) setup. This section also shows that aggregate school-level PhD student research output scales linearly with graduate program enrollment. Section \ref{sec:conclusion} summarizes our findings and concludes.

\section{Data and Descriptive Statistics}

\subsection{Data, Definitions, and Descriptive Statistics}

Our database starts with economics dissertations listed in the \emph{Journal of Economic Literature} and indexed in the ProQuest Dissertations and Theses Global database (formerly Dissertation Abstracts), augmented with information from individual schools and a few other sources detailed in the data appendix. We used ProQuest's ``research field'' classifier to identify economics dissertations.  The initial sample includes nearly 10,000 PhD graduates who completed their doctoral degrees between 1989 and 2023. Much of our analysis is limited to graduates from 1994-2017, allowing 6 years of post-PhD follow-up for the most recent cohort. The 1994 start was chosen in view of what appears to be reduced coverage of graduates from some schools in earlier years. 

\subsubsection*{Schools and Departments}

ProQuest identifies new degree-granting institutions (which we refer to as schools) more reliably than graduates' departments within schools. We were able to classify many students into departments, however, using a combination of ProQuest thesis PDFs and information supplied by economics departments at six of our eight schools.  Much of our analysis combines (and sometimes distinguishes) two groups: identifiable economics department graduates and graduates completing economics-related theses while earning degrees in economics-adjacent departments and programs. We refer to the combined sample as containing graduates of \emph{economics and related programs}.  Students appearing in department-provided spreadsheets are classified as graduating from an economics department.  Otherwise, classification into departments within schools relies on machine-reading of thesis PDFs. The data appendix details this classification process.

\begin{figure}[h!]
	\centering
	\caption{Economics and Related Program Cohort Sizes}
	\includegraphics[width=\linewidth]{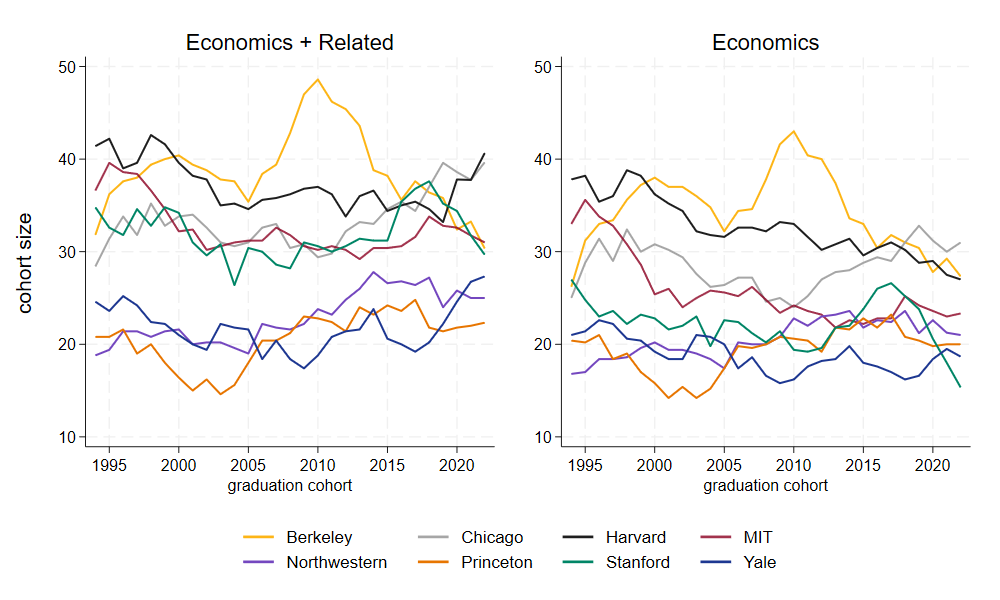}
	\label{fig: size 0}
  	\begin{flushleft}
		\noindent\scriptsize{\emph{Notes}:  The left panel shows graduation cohort size for students identified as earning either economics department degrees or related department or program degrees with affiliation determined as described in the data appendix.  The right panel shows graduation cohort size for students identified as economics department graduates. Related departments and programs include Finance, Management, Business, Accounting, Marketing, and Operations Research. A few students with no department indicated on thesis cover pages or for whom no thesis was available for download are included in the economics + related sample if one of their advisors advised at least one student identified as an economics department graduate. The figure plots five-year moving averages starting in 1994, with the first years smoothed using data back to 1989.}  
	\end{flushleft}
\end{figure}

As can be seen in the right panel of Figure \ref{fig: size 0}, economics and related-program PhD cohort size ranges from a high of over 40 for Berkeley in some years to 20 or fewer for Princeton and Yale. Average cohort sizes by school for sample periods underpinning our analysis appear in Table \ref{t:gradcounts}. The addition of graduates from economics-related programs increases economics and related cohort size much more for some schools than for others.  Specifically, economics-adjacent programs produce a larger share of graduating cohorts from Berkeley, Chicago, Harvard, MIT, and Stanford than cohorts graduating from Northwestern, Princeton, and Yale. Economics programs at the super elites--Harvard and MIT--trend downwards throughout the sample period.

\subsubsection*{Publication Data}

The sample of graduates is augmented with information on publications drawn from Econlit, an AEA database that indexes nearly 2000 economics-related journals. We focus on a shorter list of 137 journals classified as the most cited by articles published in trunk journals for economics and related disciplines plus a few other influential journals.  This ``Deep Impact'' (DI) list comes from an analysis reported in \cite{angrist2020inside}. The economics trunk journal is the \emph{American Economic Review}. We also look at publications in top-6 (T6) economics journals, defined as the usual top-5 plus the \emph{Review of Economics and Statistics}, which was once seen as roughly comparable to the top-5.\footnote{The list used here adds new AEA journals not on the DI list and a relatively new Econometric Society journal, Quantitative Economics.} 

Research activity--defined by publications--varies considerable by school and over time. Column 3 in Table \ref{t:gradcounts} shows that roughly 38-58\% of 1994-2017 graduates from economics departments published at least one DI paper in the six years following degree receipt. As can be seen in column 6, research activity rates are lower for graduates of economics-related programs.

\section{Research Output}\label{sec:researchoutput}

The advisors and graduates in our data account for a substantial slice of academic economics journal output, especially at more selective and more widely-read outlets. Appendix Figure \ref{fig: Intro_2024_v2} documents this by plotting advisor and student publication shares for various journal tiers. In the extensive set of all EconLit journals, graduates from economics and related programs and their advisors each account for roughly 5\% of articles on average over time, with the share authored by the former rising and the share authored by the latter falling through the sample period. Publication shares are markedly higher in more influential outlets. Advisors and advisees contribute over 20\% each of DI journal articles in peak years. Moreover, advisors' and advisees' research shares each peak at around half of publications in T6 outlets. These patterns underscore the outsized impact elite PhD program graduates have on academic economics research.  

\begin{figure}[h!]
	\centering
	\caption{Annual Activity Profiles: EconLit, Deep Impact, and Top 6}
	\includegraphics[width=\linewidth]{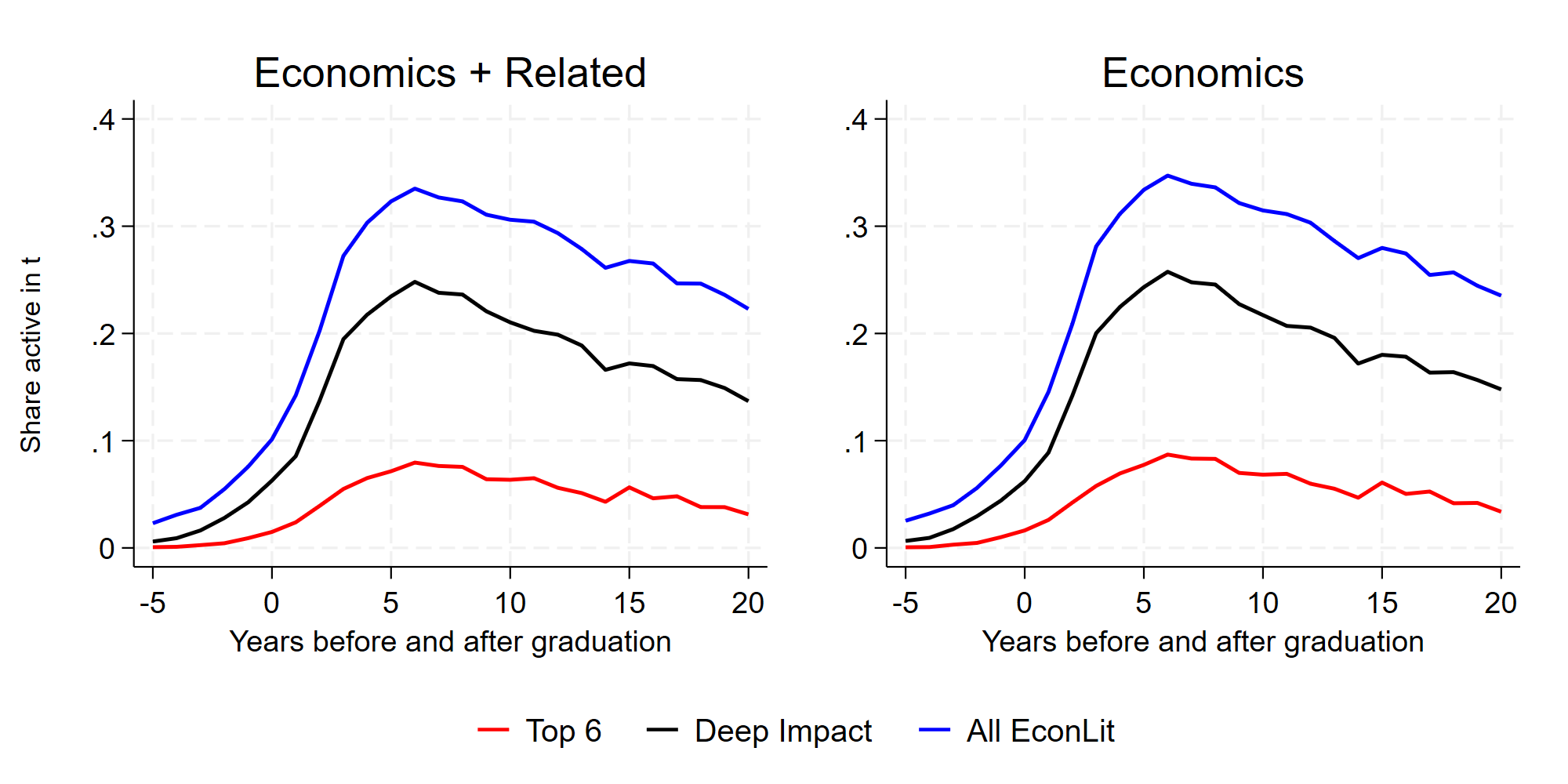}
	\label{fig: Active_combined}
        \begin{flushleft}
		\noindent\scriptsize{\emph{Notes}: This figure plots the share of students that have at least one publication in year $t-c$, where $c$ is graduation year, for three nested journal lists. Data are for economics and related program graduates on the left and for identifiable economics department graduates on the right. Data for 1994-2017 graduates.}
  
	\end{flushleft}
\end{figure}

Research activity slows early in most graduates' careers, a trend reflected in the annual activity profiles plotted in Figure \ref{fig: Active_combined}. Specifically, the figure plots the proportion of research-graduates with one or more publications in each year before and after graduation. Among graduates of economics and related programs, activity rates increase in the four years after degree completion, peaking six years after at about 8\%, 25\%, and 34\% for T6, DI, and all EconLit journals. Activity declines thereafter, a trend also documented in \cite{brogaard2018economists}. Perhaps surprisingly, annual activity profiles of (identifiable) economics program graduates, shown in the right-hand panel of the figure, are only a little above those for the broader sample that includes economics department and related program graduates.  

\begin{figure}[h!]
	\centering
	\caption{Annual Activity Profiles by School}
	\includegraphics[width=\linewidth]{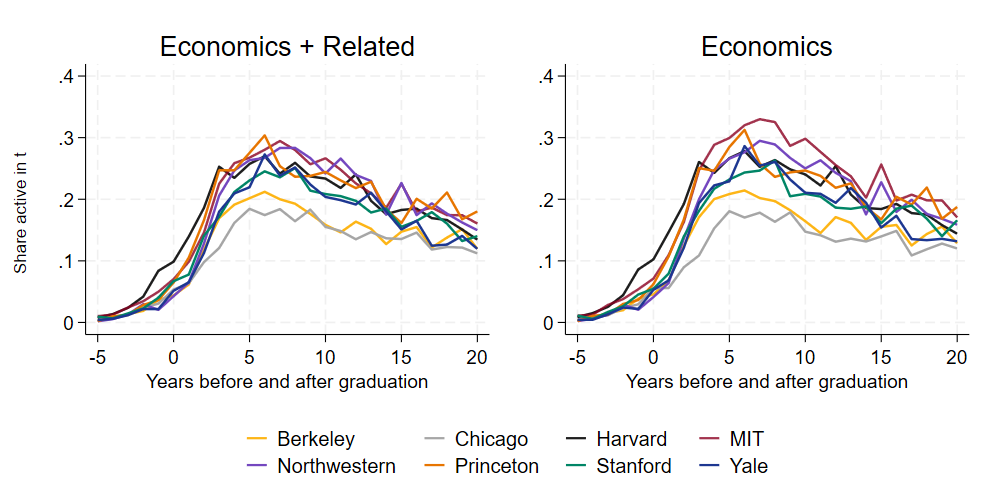}
	\label{fig: active 4}
          \begin{flushleft}
		\noindent\scriptsize{\emph{Notes}: This figure plots the share of students that have at least one Deep Impact publication in year $t-c$, where $c$ is graduation year, separately by school. Data are for economics and related program graduates on the left and for identifiable economics department graduates on the right. Data for 1994-2017 graduates.}
  
	\end{flushleft}
\end{figure}

A comparison of activity profiles by school, presented in Figure \ref{fig: active 4}, shows profiles that align with widely-held views of program prestige. Activity rates for Berkeley and Chicago graduates peak lower than do activity rates at other schools. MIT and Harvard graduates are consistently among the most active. But Northwestern and Princeton graduates' activity rates mostly track and sometimes exceed those of the Cambridge schools, while Stanford and Yale graduates are mostly in the middle.  It's noteworthy that 15 years post-PhD, activity gaps by \emph{alma mater} are much diminished. 

\begin{figure}[h!]
	\centering
	\caption{Cumulative Publication Profiles}
	\includegraphics[width=\linewidth]{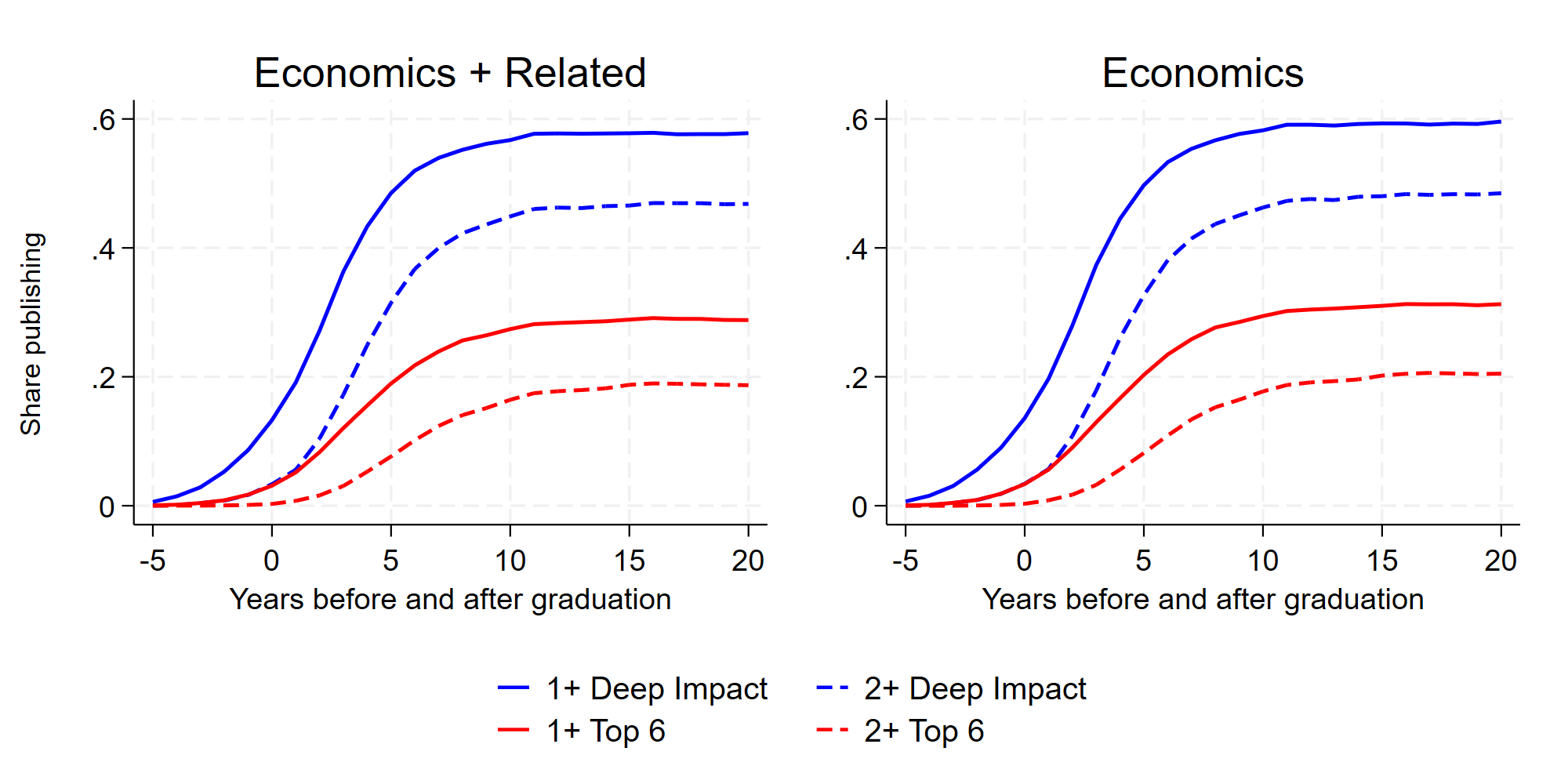}
	\label{fig: cumulative}
        \begin{flushleft}
		\noindent\scriptsize{\emph{Notes}: This figure plots the share ever-publishing as of year $t-c$, where $c$ is graduation year, for publication types and levels indicated in the legend.  Data are for economics and related program graduates on the left and for identifiable economics department graduates on the right. Data for 1994-2017 graduates.}
	\end{flushleft}
\end{figure}

Many graduates fail to produce a single publication, a finding reported earlier in \cite{conley2014research}. Figure \ref{fig: cumulative} shows something similar for our graduates. Among graduates of economics and related programs, fewer than 60\% ever place a paper in a DI journal; only around 45\% place two. These long-run success rates are only a little higher among economics department graduates.  T6 publications are rare: only 30\% of graduates place a paper in a top-6 outlet and only around 20\% manage two in T6. The likelihood of a T6 publication flattens a little sooner than does the odds of placing a paper in DI journals, again with slightly higher success rates for economics department graduates. 

\begin{figure}[h!]
	\centering
	\caption{Cumulative Publication Profiles (1+ Pubs), by School}
	\includegraphics[width=\linewidth]{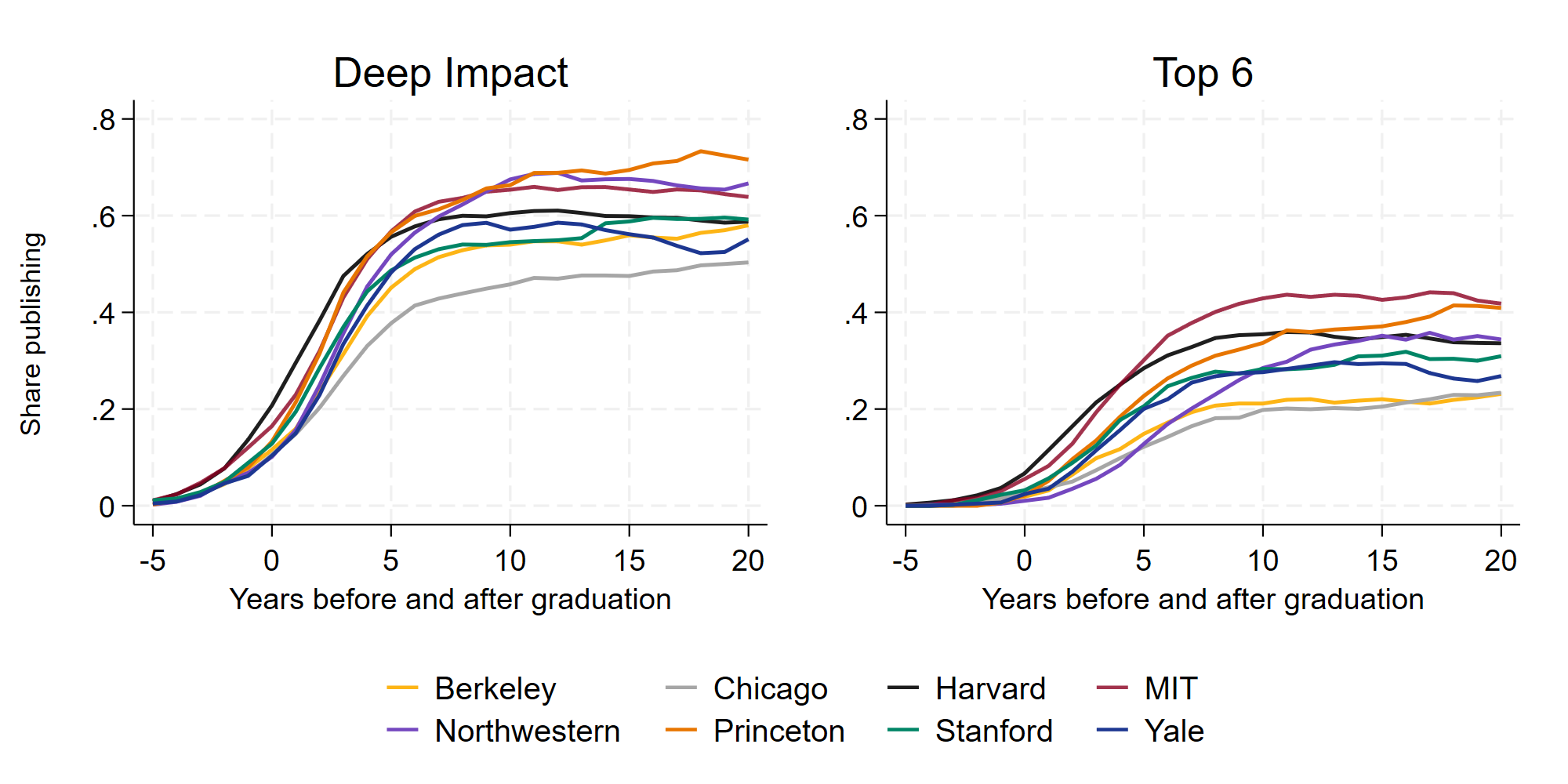}
	\label{fig: cumulative_schools}
        \begin{flushleft}
		\noindent\scriptsize{\emph{Notes}: This figure plots the share publishing (1+ pubs) in Deep Impact journals (left) and Top 6 journals (right) as of year $t-c$, where $c$ is graduation year, separately by school. Data for 1994-2017 economics+related program graduates.}
  
	\end{flushleft}
\end{figure}

Cumulative publication rates also vary markedly across schools. As can be seen in Figure \ref{fig: cumulative_schools}, cumulative DI publication rates approach 70\% for Princeton graduates, while leveling off under 50\% for Chicago graduates. Princeton and Northwestern graduates' research output tracks that of Cambridge alumni initially, eventually pulling ahead. The right side of this figure shows that MIT graduates are the most likely to place in T6 journals, with Princeton graduates a close second in the long run. Berkeley and Chicago graduates are least likely to place in a T6 outlet, a gap that persists. But the general picture is remarkably similar. Most graduates make their mark, if any, in the first 6 years after graduation.  Perhaps not coincidentally, in their 7th year of academic employment, many academics are granted a lifetime employment contract based in large part on their initial scholarship.

In the 35 years covered by our sample, elite PhD programs have grown more selective and economics doctoral student funding has grown more generous. Have more selective admissions and higher graduate-education spending yielded higher research output? Cross-cohort trends in research success are captured by a Poisson regression of DI and T6 publication counts on cohort and school effects that can be written:
\begin{equation}
E[R_{isc}|c(i), s(i)] = e^{\delta_{s(i)}+\gamma_{c(i)}},\label{eq:poisson}
\end{equation}
where dependent variable $R_{isc}$ is the year $c+1$ to $c+6$ research output of graduate $i$ from school $s(i)$ in cohort $c(i)$, parameters $\delta_s$ are school effects and parameters $\gamma_c$ are cohort effects. Cohort effects in expression \eqref{eq:poisson} capture differences in research output relative to the reference year, controlling for cross-cohort changes in the graduate distribution over schools.\footnote{Appendix Figure \ref{fig: Fraction_schools_degreeyear} plots research activity by cohort and school.}  

\begin{figure}[h!]
	\centering
	\caption{Cohort Effects in Activity Rates}
	\includegraphics[scale=.9]{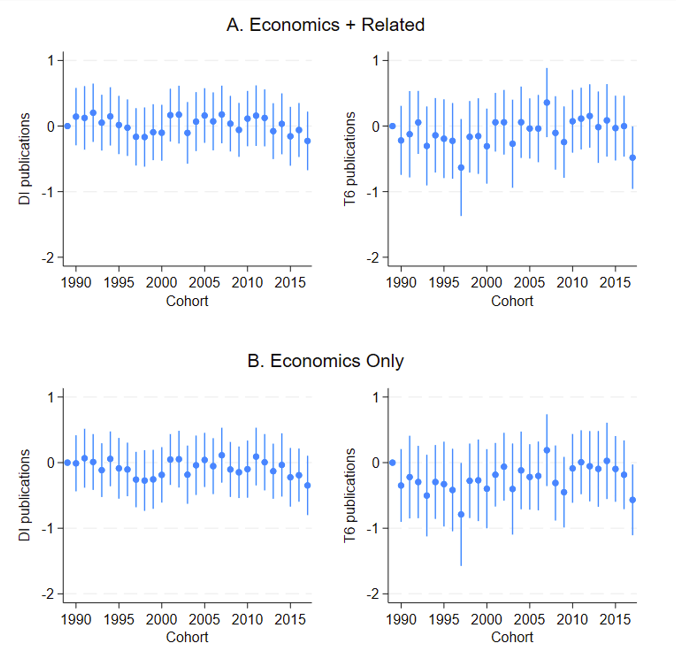}
	\label{fig: Cohorts_activity_econ_related}
    	\begin{flushleft}
 	\noindent\scriptsize{\emph{Notes}: This figure reports estimated cohort effects from a student-level Poisson regression of Deep Impact and Top 6 publication counts for years c+1 to c+6 on cohort and school effects. The reference year is 1989 and the sample starts with the 1989 cohort. Results for Deep Impact are on the left; results for Top 6 are on the right. Estimates in Panel A use the sample of economics and related program graduates (dependent variable means are 1.59 for Deep Impact publication counts and 0.39 for Top 6 publication counts). Panel B reports estimates for identifiable economics department graduates only (dependent variable means are 1.65 for Deep Impact publication counts and 0.42 for Top 6 publications). Standard errors are clustered on school-by-year.}
	\end{flushleft}
\end{figure}

Increased program selectivity and spending notwithstanding, estimated cohort effects suggest graduate research productivity has changed little since the 1989 cohort. This is apparent in Panel A in Figure \ref{fig: Cohorts_activity_econ_related}, which plots estimates of $\gamma_c$ for DI and T6 publications; 1989 is the reference cohort with an effect of zero. For DI publications, estimates hover around zero (meaning no difference from 1989). For T6 publications, cohort effects range from mostly negative before the early 2000s to mostly small and positive after 2007. This modest increase may reflect longer degree completion times, allowing early-career scholars to learn how to clear ever-higher bars for T6 acceptance.\footnote{\cite{card2013nine} document declining acceptance rates at top-5 journals from 1990-2012, while also showing a roughly proportional rise in number of authors per paper published. \cite{ellison2002slowdown} reports increasing review times at T6 journals.} Panel B of the figure reports estimates computed using the sample of identifiable economics department graduates only. Estimated cohort effects for economics graduates differ little from the estimates for the broader sample plotted in Panel A.

\subsection{Gender Gaps}\label{sec:gender}

\begin{figure}[h!]
    \centering
    	\caption{Graduate and Advisor Share Female}
    \includegraphics[width=0.9\linewidth]{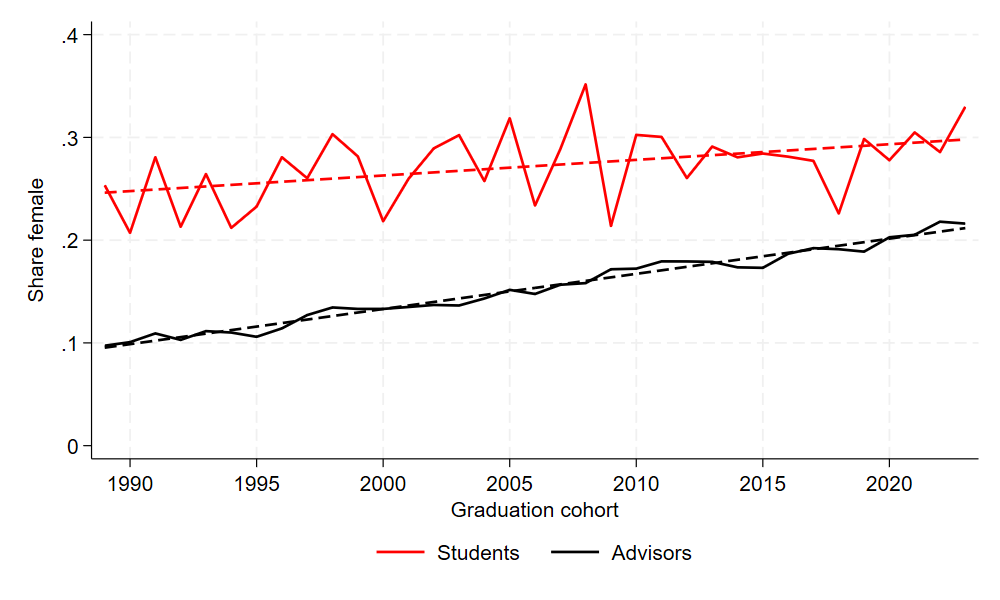}
    \label{fig: female shares}
           \begin{flushleft}
    \noindent\scriptsize{\emph{Notes}: The red line plots the graduate share female by graduation cohort; the black line plots the share female among advisors. Gender is coded from first-name gender frequencies in Social Security records, as described in the data appendix. Advisor data is limited to years of active advising, defined as the period between an advisor's first and last advisee graduation year. Data are for the economics + related sample of 1989-2023 graduates and their advisors. }
    \end{flushleft}
    
\end{figure}

We conclude this descriptive analysis of graduate research productivity with a brief look at gender gaps in publications. As can be seen in Figure \ref{fig: female shares}, fewer advisors than students are female, a pattern that likely reflects the fact that the advising load is concentrated among successful academic researchers, fewer of whom are female.  At the same time, the share of advisors who are female has climbed steadily from around 9\% in 1989 to roughly 22\% in 2023. Although more volatile than advisor share female, the share of PhD graduates who are female has also trended upwards, and now exceeds 30 \%.  

\begin{figure}[h!]
	\centering
	\caption{Gender Gaps in Graduate Research Annual Activity Profiles}
	\includegraphics[width=\linewidth]{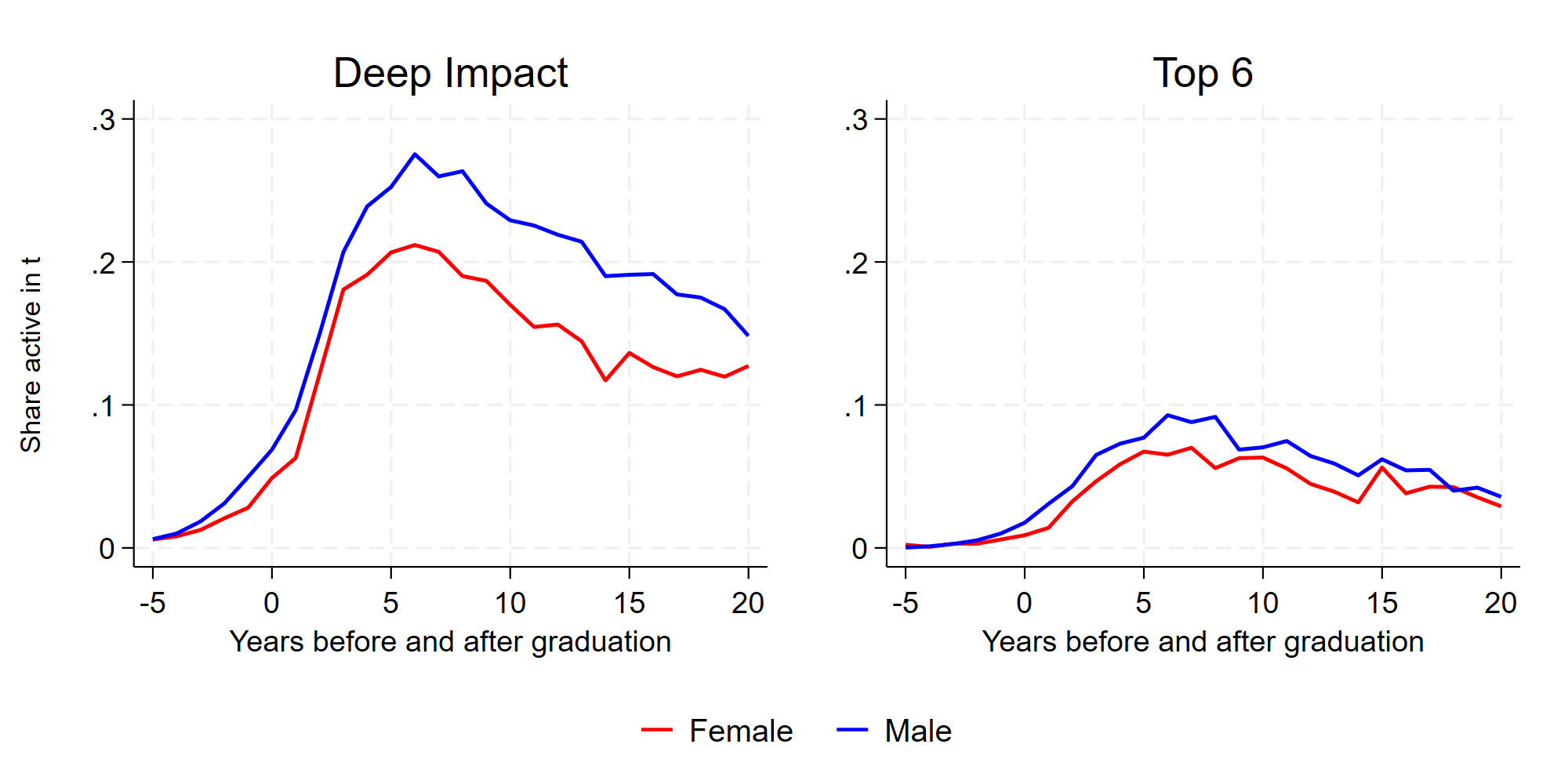}
	\label{fig: activity gaps by gender}
           \begin{flushleft}
		\noindent\scriptsize{\emph{Notes}: See notes to Figure \ref{fig: female shares}. This figure plots the share of male and female students with at least one publication in the Deep Impact (left panel) or Top 6 (right panel) journal list in year $t-c$, where $c$ is graduation year. Appendix Figure \ref{fig: active 2 counts} plots publication counts in the same format. Data are for the economics and related sample of of 1994-2017 graduates.} 
      
	\end{flushleft}
\end{figure}

Research activity profiles by time since degree, plotted by gender in Figure \ref{fig: activity gaps by gender} show male and female graduates to be similarly active in the first few years post-PhD.  By year five, however, female DI activity rates have crested while male DI activity rates continue to climb.  This leads to a gap in activity rates that begins to close only 15 years after graduation. Consistent with much lower T6 activity, gender gaps in T6 publication rates are smaller than those for DI, and appear to close sooner than those for DI publications. As a proportion of corresponding male activity rates, however, the largest gaps (observed between post-PhD years 5-10) are similar for DI and T6 publications.\footnote{Proportional gaps are $0.23$ for DI and $0.20$ for T6.}

\begin{figure}[h!]
	\centering
    \caption{Annual Activity Profiles by Graduate Student and Advisor Gender}	\includegraphics[width=0.8\linewidth]{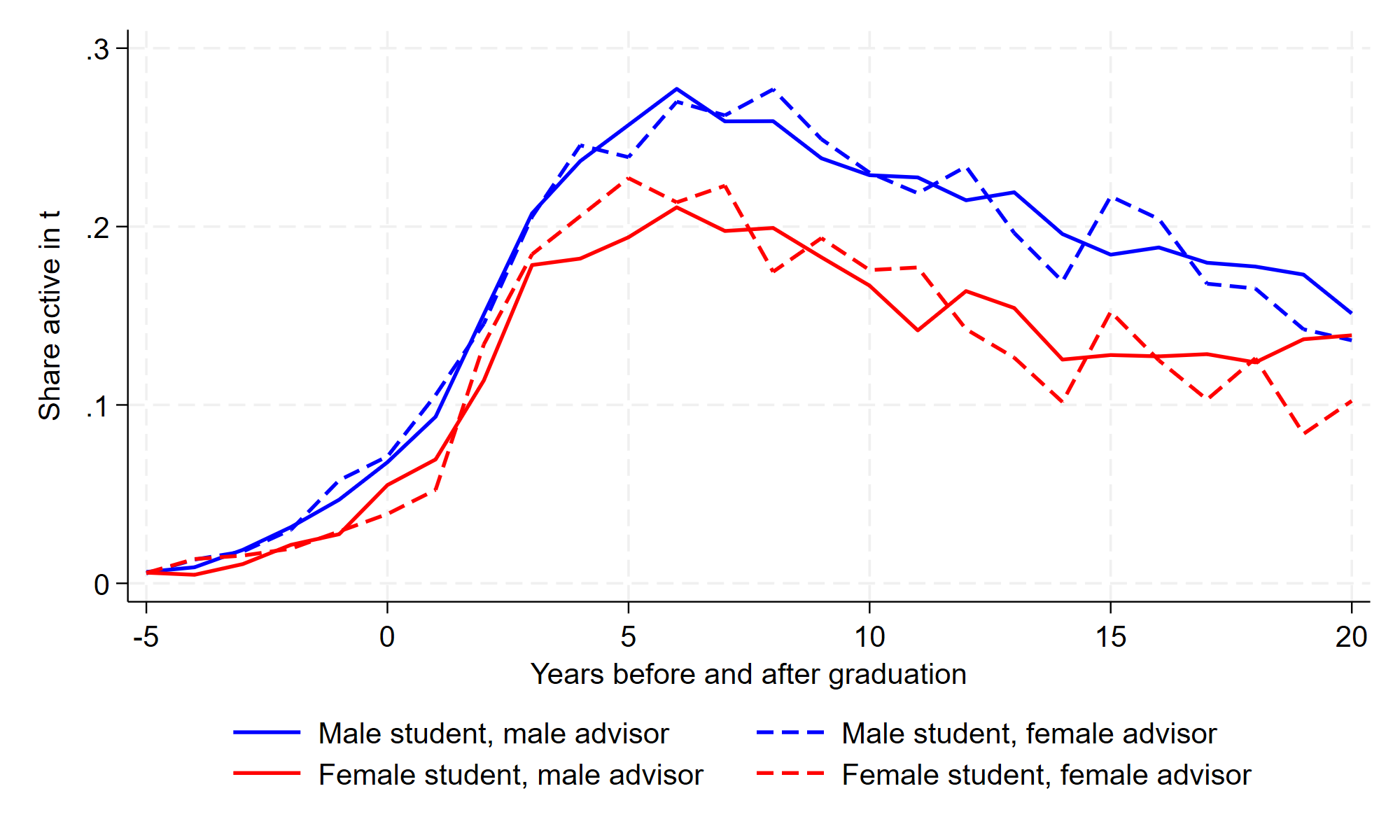}
	\label{fig: active 3}
           \begin{flushleft}
		\noindent\scriptsize{\emph{Notes}:  See notes to Figure \ref{fig: female shares}. This figure plots the share of students with at least one publication in the Deep Impact journal list by advisee gender and advisor gender in year $t-c$, where $c$ is graduation year. Graduates marked as having male advisors were advised by all-male advising teams. Graduates marked as having female advisors had at least one female advisor.  Appendix Figure \ref{fig: active 3 counts} plots publication counts in the same format.  Data are for the economics + related sample of 1994-2017 graduates.}
	\end{flushleft}
\end{figure}

\begin{figure}[h!]
	\centering
	\caption{Gender Gaps in Annual Research Productivity by Cohort}
	\includegraphics[width=\linewidth]{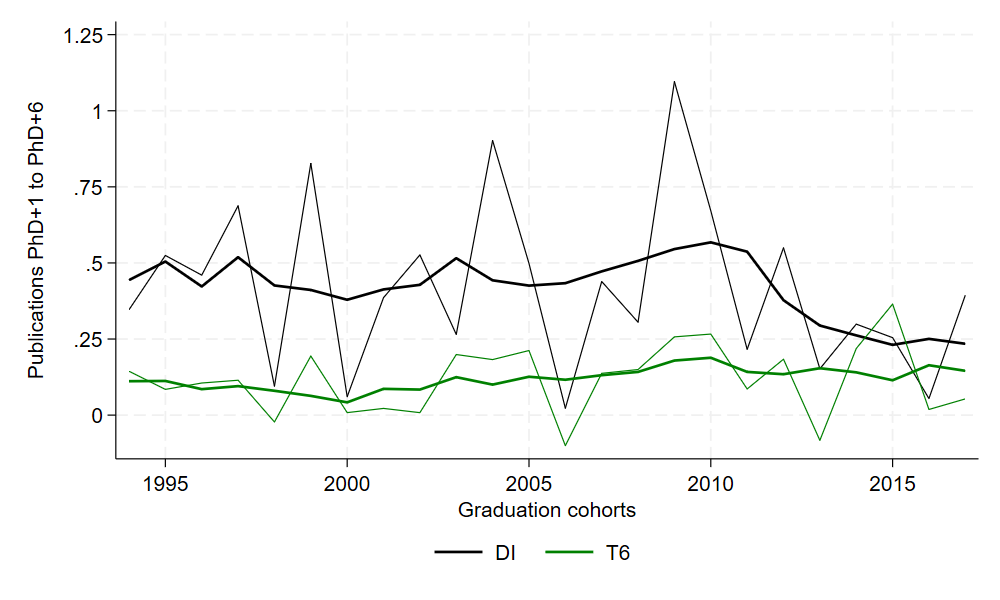}
	\label{fig: cohort gender gap}
 \begin{flushleft} 
 		\noindent\scriptsize{\emph{Notes}:  See notes to Figure \ref{fig: female shares}. This figure plots gender gaps in the number of Deep Impact and Top 6 publications by cohort. For graduates in cohort $c$, publications are counted in years $c+1$ to $c+6$. The figure shows the difference in average productivity between male and female graduates. Thicker lines show a 5-year moving average. Data are for the economics + related sample of 1994-2017 graduates.}
   \end{flushleft}
\end{figure}

Gender gaps in activity rates appear to be unrelated to advisor gender. Figure \ref{fig: active 3} plots activity profiles similar to those plotted in Figure \ref{fig: activity gaps by gender}, separately by advisor as well as advisee gender. Activity rates of female and male students evolve similarly regardless of advisor gender. In particular, activity rates of female students with at least one female advisor fall relative to activity rates for men who were advised by either men or women. Interestingly, however, the gender gap in DI publication rates is lower on average for recent cohorts, a pattern suggested by the smoothed DI gap in the Figure \ref{fig: cohort gender gap}.  The gender gap in T6 publications, by contrast, is reasonably stable across cohorts.\footnote{\cite{iaria2024gender} likewise highlights lower publication rates for female scientists. \cite{card2020referees} analyzes gender differences in refereeing and citations for papers reviewed at four economic journals, arguing that referees judge male- and female-authored papers similarly. \cite{Kjelsrud2024gender} argues that the presence of female faculty improves female economics graduate student performance post-PhD while reducing post-Phd outcomes for male students. Figure \ref{fig: active 3}  seems inconsistent with this.} 

Persistent gender gaps in research success and changes in gender composition across cohorts lead us to control for student gender in regression models that aim to predict student research success. Advisor gender and student-advisor gender matching appear less important.  We focus, therefore, on advisor characteristics that seem directly related to research and advising success.\footnote{Appendix Figures  \ref{fig: active 2 counts} and \ref{fig: active 3 counts} do versions of Figures \ref{fig: activity gaps by gender} and \ref{fig: active 3} for publication counts; these show gender gaps similar to those in activity rates.}

\section{The Advising Relationship}\label{sec:advising}

\begin{figure}[h!]
	\centering
	\caption{The Advising Load Distribution}	
	\includegraphics[width=\linewidth]{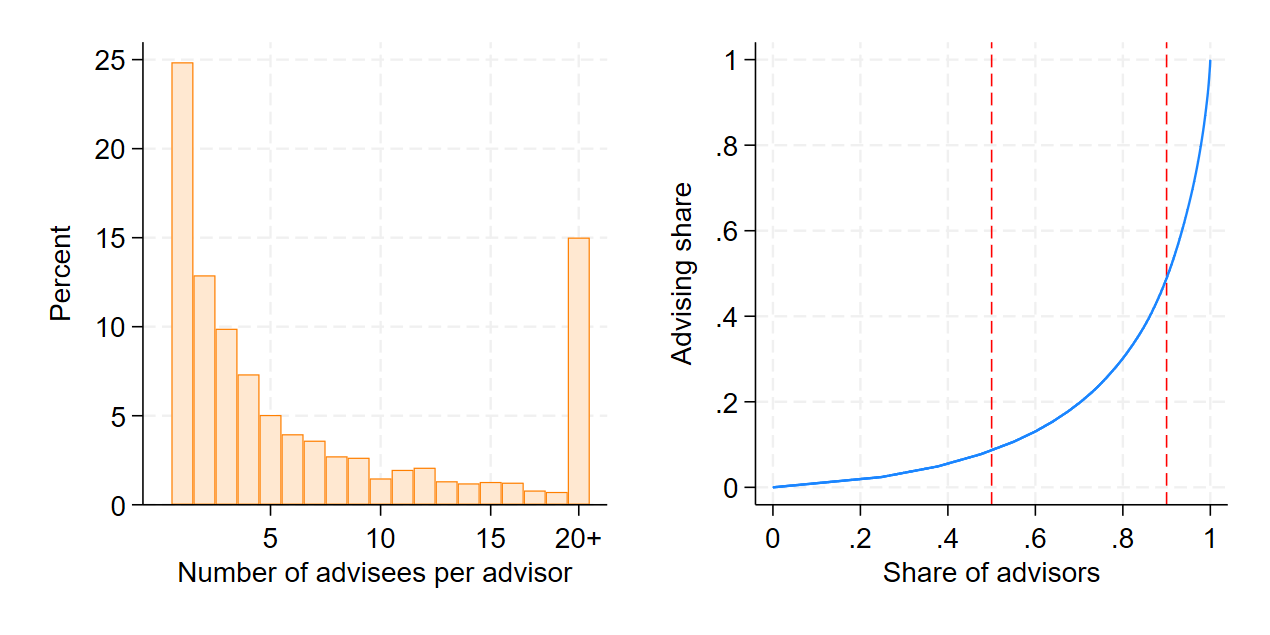}
	\label{fig: load}
	\begin{flushleft}
		\noindent\scriptsize{\emph{Notes}: The left panel shows the histogram of number of advisees for advisors who advised an economics+related program graduate and were affiliated with one or more of the eight schools in our sample (advisors necessarily have at least 1 advisee who graduated from sample schools but need not be affiliated with a sample school or have advised an economics+related program graduate). Each advisor contributes one observation. The right panel shows the advising Lorenz curve: this orders advisors on the x-axis by number of advisees, with the cumulative share of advisees advised up to this point plotted on the y-axis. Red lines mark median and upper-decile advisors. Data are for 2499 advisors of economics + related program students who graduated 1989-2023; advisors in this sample have at least one EconLit publication from which an affiliation can be gleaned.} 
	\end{flushleft}
\end{figure}

The distribution of graduate advising is highly skewed, with a minority of advisors advising a large share of graduates. This is documented in Figure \ref{fig: load}, the left panel of which shows the histogram of number of advisees for the 2499 advisors in our data who advised at least one student in the sample of 1989-2023 graduates from economics and related programs and who were affiliated with one of our eight schools.\footnote{Advisors necessarily have at least 1 advisee who graduated from sample schools, but need not be \emph{affiliated} with a sample school or have advised an economics+related program graduate. Advisor affiliations are taken from advisor publications in EconLit; for 1171 with no EconLit publications, this information is missing. And many of the roughly 4800 advisors in our data have an affiliation outside our eight-school sample.} Roughly a quarter of advisors have only one advisee in the relevant cohorts, while the busiest 15\% of advisors advised 20 or more PhD students. The Lorenz curve shown in the right panel of the figure highlights this concentration further. The least prolific 50\% of advisors (in terms of advising load) advised fewer than 10\% of graduates (indicated by a red line at $0.5$ on the x-axis). At the other end of the distribution, the 10\% of advisors who are busiest account for roughly half of the advising relationships in our sample.

\subsection{Advisor Attributes and Student Success}

Advisors vary greatly both in their own research output and in the number of students they advise.  Do the most successful faculty researchers advise the most successful PhD students?  Is past advising success--both in terms of numbers and in terms of student publications--a predictor of future advisee performance?  We construct three sorts of variables to quantify these aspects of the advising relationship.  

Advisor research productivity for a cohort graduating in year $c$ is characterized here by an advisor's publication record in the five years preceding $c$. Specifically, \emph{advisor research}, denoted $AR_i$, counts advisors' DI publications for each 5-year window ($c-5$ to $c-1$), averaged over advisors for each student for up to six advisors (roughly 80\% of students have 3 or fewer advisors). Let $A_{ic}$ denote the set of advisors, indexed by $j$, who advised PhD graduate $i$ finishing in cohort $c$.  Then $AR_i$ is defined by:
\begin{equation}\label{eq:defineAR}
AR_i \equiv \frac{1}{|A_{ic}|}\sum_{\{j \in A_{ic}\}} R_{jc} \ ,
\end{equation}
where $|A_{ic}|$ is the number of advisors to $i$ and $R_{jc}$ is the $jth$ advisor's DI publication count from $c-5$ to $c-1$. Similarly, our measure of \emph{advising load} averages the number of a cohort-$c$ graduate's advisors' advisees finishing in $c-5$ to $c-1$, over up to six advisors for a given student.  This quantity, denoted, $AL_i$, can be written
\begin{equation}\label{eq:defineAL}
AL_i \equiv \frac{1}{|A_{ic}|}\sum_{\{j \in A_{ic}\}} L_{jc} \ , 
\end{equation}
where $L_{jc}$ counts advisor $j$'s advisees in the relevant cohorts. Note that, for a given graduate, both $AR_i$ and $AL_i$ are averages over the advisor team. 

Finally, \emph{past student success} averages the number of DI publications by an advising team's past students. For cohort-$c$ graduate $i$ advised by advisors indexed by $j \in A_{ic}$, this variable averages DI publications by these advisors' advisees in the six years since the advisees graduated, looking at advisor-$j$ advisees who finished in years $c-5$ to $c-1$. Specifically, let $S_j^{(i)c}$ denote the set of students indexed by $t \ne i$, advised by $j$ who graduated in cohorts $d \in [c-5, c-1]$. Past student success, denoted, $PS_i$, is defined as
\begin{equation}
\label{eq:definePS}
PS_i \equiv \frac{1}{|A_{ic}|}\sum_{\{j \in A_{ic}\}} \frac{1}{|S_j^{(i)c}|}\sum_{\{t \in S_j^{(i)c}\}} R_{td}, 
\end{equation}
where $R_{td}$ counts DI publications by student $t$ who graduated in cohort $d$ in years $d+1$ to $d+6$. 

Like research success, advising success is rare. We therefore look at upper-tail measures of success as well as advisor averages. For student $i$ who graduated in cohort $c$, three \emph{super} advising dummy variables indicate whether at least one of $i$'s advisors was in the top 10\% of advisors among those who advised students who graduated from $c-5$ to $c-1$.  For example, the \emph{super advising load} dummy indicates graduates in cohort $c$ with at least one advisor in the top 10\% of advisors measured by number of advisees graduating in $c-1$ to $c-5$. This amounts to having one advisor $j \in A_{ic}$ with $L_{jc}$ in the top 10\% of the distribution of $L_{jc}$ for the relevant set of cohorts as a group. Super advisor research and super past student success are defined similary. It's enough to have one super advisor for a student-level super dummy to switch on. A parallel set of three \emph{duper} variables indicate graduates with at least one advisor among the 5\% most prolific when ranked by advisor research, advising load, and past student success.\footnote{Advisors are defined as super using the universe of economics-related PhDs, not limited to advisors of PhDs from economics departments and related programs. The super dummy for past student success is coded as follows: for each advisor, advisees graduating $d \in [c-5, c-1]$ are identified. DI publications for this group in years $d+1$ to $d+6$ are summed and divided by the number of advisees graduating in this period. This quantity (the inner summation in expression \eqref{eq:definePS}) gives an advisor's average past student success looking at their advisees from the last five years. A student is advised by a super advisor if one of their advisors is among the top 10\% of advisors when ranked by average past student success.}

\begin{figure}[h!]
	\centering
	\caption{Super Advisors by Cohort and School}
	\includegraphics[width=0.6\linewidth]{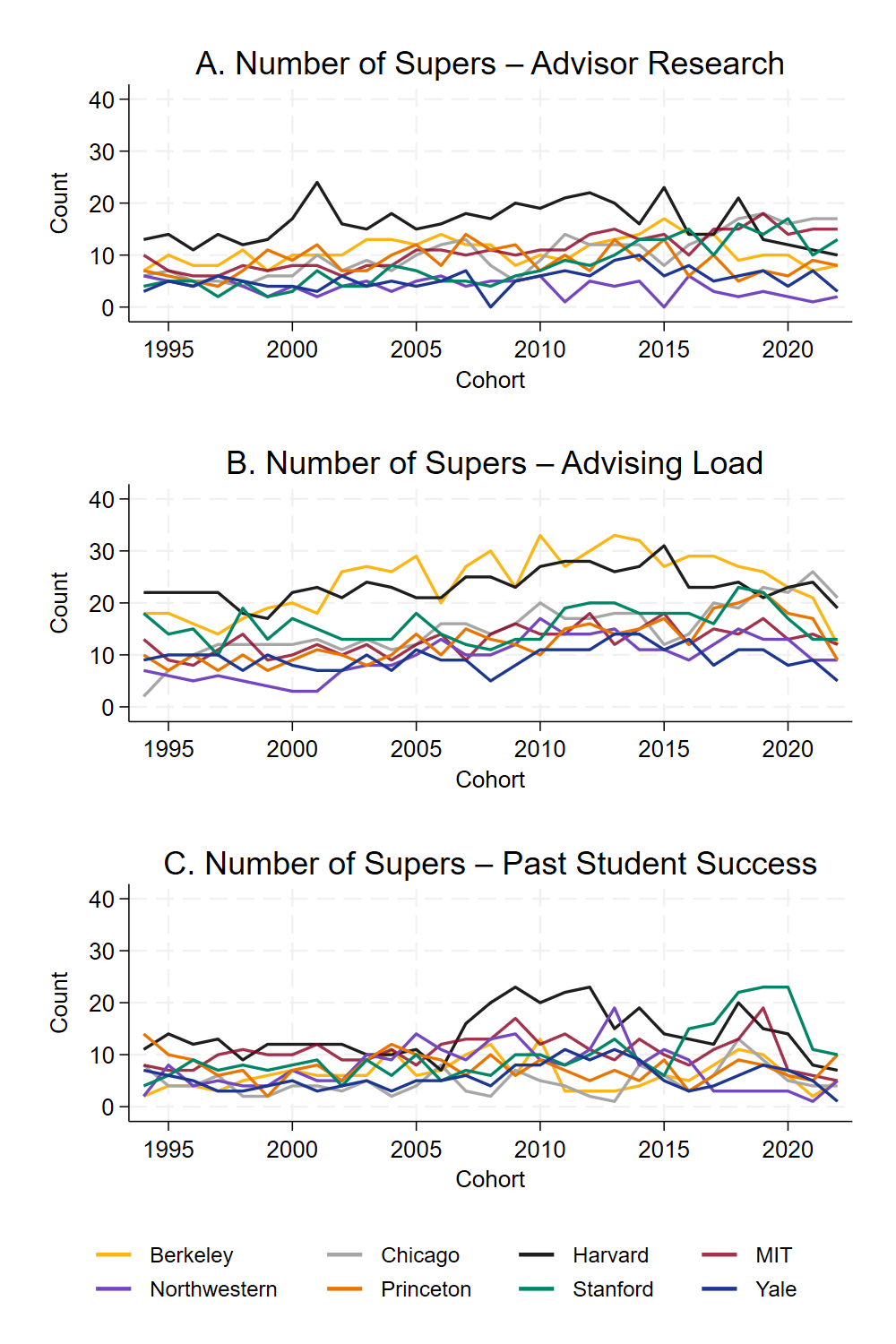}
	\label{fig: rollout_supers_without_allways}
    	\begin{flushleft}
 	\noindent\scriptsize{\emph{Notes}: This figure plots the number of super advisors affiliated with a given school in a given year that advised at least one student in cohort $c$.  Super advisors are classified by cohort using advisor attributes in the five years preceding the cohort's graduation year. Successful-research advisors have published many Deep Impact articles; advising-loaded advisors have advised many PhDs; successful-student advisors have advised graduates who published many Deep Impact articles in the six years following their graduation. Super advisors are in the top 10\% of advisors in these categories, defined separately for each distribution. Advisor distributions are for the universe of economics PhDs, not limited to graduates of economics+related programs. }
	\end{flushleft}
\end{figure}

Figure \ref{fig: rollout_supers_without_allways} tracks the distribution of super advisors over time by school. Consistent with the traditional view of Harvard as employing many highly successful senior scholars, Panel A in this figure shows Harvard advisors enjoy standout research success, while Northwestern and Yale advisors are the least successful.\footnote{Harvard advising outcomes are also affected by school size, with substantial economics-adjacent graduate enrollment at Harvard's Kennedy School, Business School, and Graduate School of Education.} Interestingly, beginning in the early 2000s, Berkeley moves ahead with the largest number of super advisors in terms of advising load. Perhaps not coincidentally, this increase follows legendary advisor David Card's move to Berkeley. Not only is Card a super advisor, he attracted top younger advisors to the Berkeley faculty. The bottom panel of Figure \ref{fig: rollout_supers_without_allways} offers an interesting counterpoint to the top two panels: student research success shows less dispersion across schools than do advising load and advisor research success.

A student-level regression connects graduates' post-PhD research success with advisor attributes, controlling for cohort and school effects. Let $R_{isc}$ denote publications by student $i$ graduating from school $s(i)$ in cohort $c(i)$, including zeros. Specifically, the impact of advisor attributes on $R_{isc}$ is estimated using the following regression model:
\begin{equation}\label{eq:studentmodel}
R_{isc} =  \alpha'W_i + \tau'D_i  + \beta N_i + \gamma_{c(i)} + \delta_{s(i)} + \varepsilon_{isc}, 
\end{equation}
where $D_i$ is a vector of one or more of the advisor variables defined above, $N_i$ is the number of $i$'s advisors, and $\delta_{s(i)}$ and $\gamma_{c(i)}$ are the relevant school and cohort effects, respectively. Control for team size is motivated by the fact that, for a given graduate, average advisor productivity, advising loads, and past student success are diluted when advising teams are larger. On the other hand, the probability of having at least one advisor with upper-tail values of these variables increases with the size of the advising team. Vector $W_i$ in equation \eqref{eq:studentmodel} contains a set of student controls that includes a female dummy, a dummy for PhDs with gender unclassified, a dummy for graduates with no readable thesis PDF, and a dummy for identifiable economics department graduates (and a constant). Vector $\tau$ contains the coefficients of primary interest. We also report estimates of a Poisson analog of equation \eqref{eq:studentmodel}, where estimates of $\tau$ give the percentage change in publications attributable to $D_i$.

In models entering advisor attributes one at a time, graduates advised by advisors who publish more, advisors with more advisees, and advisors with more successful former students have greater publication success.  The first column of Panel A in Table \ref{t:studentregs} shows, for instance, that graduates see $0.12$ more DI publications, on average, when advising team research increases by one. This is a gain of roughly 7\% (the corresponding Poisson coefficient appears in column 5). As can be seen in column 3, advisor research success appears to boost advisees' T6 publications much less, though effects on advisees' T6 publications in percentage terms (reported in column 7) are larger and approach 9\%. The estimates in the second row of Panel A show that graduates advised by super researchers generate more publications. Super-research effects are $0.70$ and $0.24$ for DI and T6 levels, respectively, and $0.47$ and $0.67$ in percent.\footnote{\cite{garcia2020predicting,hilmer2009fishes} likewise find a strong association between advisor research productivity and economics advisees' post-PhD publishing success.}

The estimates in Panel B of Table \ref{t:studentregs} suggest that an advising team's advising load is (mostly) a weaker predictor of student research success than the team's own research record. Patterns here are similar to those in Panel A, however, with consistently positive effects: prolific advisors (in terms of numbers of past students) have advisees that see greater post-graduation research success. As with the estimated advisor-research effects in Panel A, effects on student output in levels are larger for DI than T6, while effects in percentage terms are larger for T6 than for DI. Super advising-loaded advisors have substantially more successful students than do advisors with fewer advisees.    

Among the three continuous advisor attributes examined in Table \ref{t:studentregs}, the estimates in Panel C of the table show that average past student success is the best predictor of current student success. Increasing an advisor team's average past student DI publications by 1, for instance, is associated with $0.43$ more current student DI publications, a 25\% increase. Increasing the average of past student DI publications is predicted to yield a 35\% increase in current student T6 publications. Again, coefficients on super dummies indicating the strongest 10\% of advisors based on their advisees' past success are even larger.  

Multivariate models that capture effects of advisor research, advising load, and past student success jointly suggest advisor research and past student success are more important drivers of current student success than advising load. This can be seen in Table \ref{t:multistudent}, which reports OLS estimates in columns 1-4 and Poisson estimates in columns 5-8, separately for DI and T6 student publications. In particular, the estimates in columns 1, 3, 5, and 7 show advising load effects close to zero and not significantly different from zero.  The corresponding estimates for advisor research and past student success are smaller than those in Table \ref{t:studentregs}, though still substantial and significantly different from zero. 

Even-numbered columns in Table \ref{t:multistudent} report results from models that add duper dummies indicating graduates advised by at least one advisor in the upper 5\% of the relevant advisor attributes distribution.  In these specifications, super and duper advising dummies are mutually exclusive, so super dummies indicate advisors with characteristics in percentiles 6-10. Estimated super advising effects remain large and relatively precisely estimated for all 3 variables, though they're smaller than the corresponding one-at-a-time estimates in Table \ref{t:studentregs}. The multivariate super advising load coefficient is the most diminished from Table \ref{t:studentregs}, though still positive and mostly at least marginally statistically significant. Duper advising coefficients are larger than the corresponding super advising coefficients, which are all positive, suggesting that student success is monotone in advisor quality as proxied by advisor research, advising load, and past student success.

\begin{figure}[h!]
	\centering
	\caption{Student Success and No. of Advisees for 236 Advising-Loaded Advisors}
	\includegraphics[width=\linewidth]{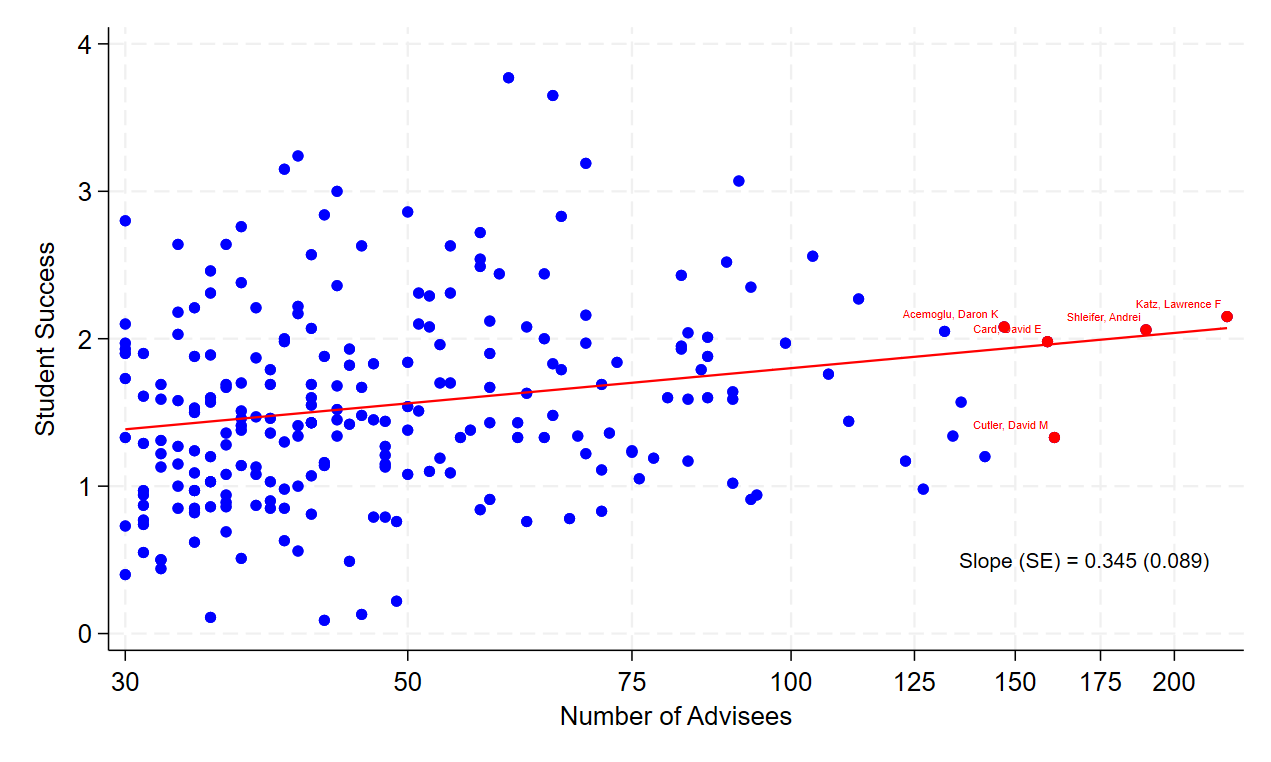}
	\label{f:tobyfig}
    	\begin{flushleft}
 	\noindent\scriptsize{\emph{Notes}: This figure plots student success for prolific advisors (who advised at least 30 graduates) against each advisor's number of advisees. Both variables are computed for graduates who earned degrees at one of the eight schools in our sample in 1989-2023. Student success averages post-graduation publications in years c + 1 to c + 6 for a graduate in cohort $c$.}
	\end{flushleft}
\end{figure}

Multivariate models generate dampened effects of advising load on student research success. Interestingly, however, the estimates in Tables \ref{t:studentregs} and \ref{t:multistudent} show no evidence of a productivity-diminishing advising burden: advisees guided by advisors carrying a heavy load do not appear to suffer when their advisors' attention is divided more finely. Figure \ref{f:tobyfig} highlights the weakly positive relationship between advising load and student research success. This figure plots average student success for all 1989-2023 graduates against the number of students advised, tabulated for the sample of advisors who advised at least 30 students between 1989-2023.\footnote{Let $S_j$ denote the set of advisor $j$'s students, indexed by $i$, who graduated 1989-2023. Overall student success in Figure \ref{f:tobyfig}, denoted $PS_j$ for advisor $j$, is defined as: 
        \begin{equation} 
        PS_j \equiv \frac{1}{|S_j|}\sum_{\{i \in S_j \} } R_{i}, 
        \end{equation}
        where $R_{i}$ counts DI publications by student $i$ who graduated in cohort $c$ in years $c+1$ to $c+6$.}
The x-axis is a log scale, so the slope of the fitted line implies that a 10\% increase in advising load is associated with $0.035$ additional post-PhD publications on average, with considerable variation around this.

The figure is also consistent with the robustly positive duper advising load coefficient estimates reported in Table \ref{t:multistudent} in suggesting an outsize role for exceptionally prolific advisors (measured by their advising loads). The five most prolific advisors (Acemoglu, Cutler, Card, Katz, and Shleifer) are labeled in the figure; these lie Northeast of the mass of points marking other advisors. Figure \ref{f:tobyfig} is complemented by Appendix Table \ref{t:league} and Appendix Figure \ref{f:rankfig}. The table lists the 189 most heavily-loaded advisors, ranked by their students' average DI publishing success. This table includes advisor rank by student T6 publications, number of advisees, and advisor rank by advising load.\footnote{The table is truncated from below at prolific advisor and Nobel laureate Robert E. Lucas, whose many advisees averaged just over 1 publication post-PhD. The figure plots data for all 236 advisors who advised at least 30 students.} Appendix Figure \ref{f:rankfig} plots advisor rank based on student T6 publications against advisor rank based on student DI publications. These measures are highly correlated, with a regression slope of $0.81$ and an $R^2$ of $0.66$. Of course, the correlations and regression estimates discussed in this section may reflect student selection or sorting as well as causal effects.  We return to this point following an examination of student-driven aspects of the advising relationship.  

\subsection{Advisor-Advisee Coauthoring and Research Affinity}

\begin{figure}[h!]
	\centering
	\caption{Early Career Advisor-Advisee and Classmate Coauthoring by Cohort}	
	\includegraphics[width=\linewidth]{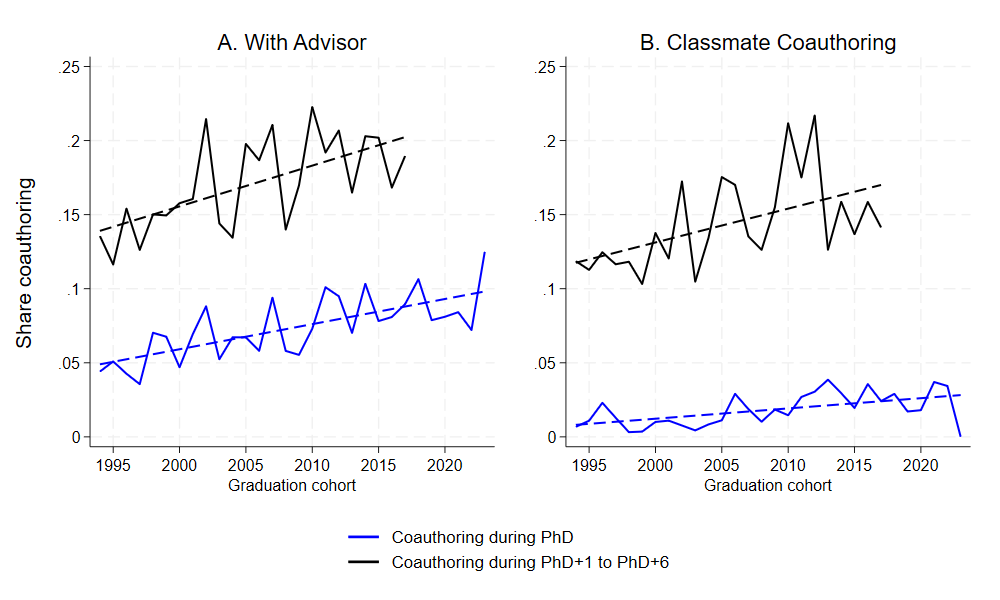}
	\label{fig: Co_mean_5years_postgrad}
	\begin{flushleft}
  	\noindent\scriptsize{\emph{Notes}: 
The left panel shows trends in student-advisor coauthoring; the right panel shows trends in PhD student coauthoring with classmates. Classmates are defined as students from the same school who graduated in the same year or within two years before or after. Dashed blue lines mark the share of students with a pre-graduation publication authored jointly with an advisor or classmate including in the graduation year. Dashed black line plots the share of students with at least one joint publication in the six years following graduation, excluding the graduation year. Coauthoring is determined from all publications indexed in EconLit for 1994-2023 cohorts.}
	\end{flushleft}
\end{figure}

Advisor-advisee coauthoring has long been common in the natural sciences.\footnote{See, for instance, \cite{azoulay2010}, which examines collaborations in the life sciences.} Mirroring the rise of empirical economic research since the early 1990s, this sort of teamwork has grown in economics publications too \citep{econevolves,jones2021teams}. Figure \ref{fig: Co_mean_5years_postgrad} documents increasing student-advisor and student-classmate coauthoring for cohorts graduating since 1994.  Specifically, the figure tracks the share of students coauthoring with either an advisor or a classmate (defined as same-school PhDs who graduated the same year or within two years before or after), before and after degree receipt. 

Unsurprisingly, coauthoring of any kind is far more common after degree receipt than before (since PhD students must publish to coauthor). For all cohorts, coauthoring with advisors is more common than coauthoring with classmates. But both sorts of coauthoring are on the rise. Among the most recent cohorts for which data appear in Figure \ref{fig: Co_mean_5years_postgrad}, advisor-advisee coauthoring rates hit 20\%, while classmate coauthoring reaches roughly 17\%.\footnote{\cite{sarsons2021gender} examines gender differences in credit attribution for joint work, finding statistically insignificant gender gaps in post-tenure publications and citations to jointly-authored research.}
 
\begin{figure}[h!]
	\centering
	\caption{Share of Theses Citing One or More Advisors by Cohort and School}	
	\includegraphics[width=\linewidth]{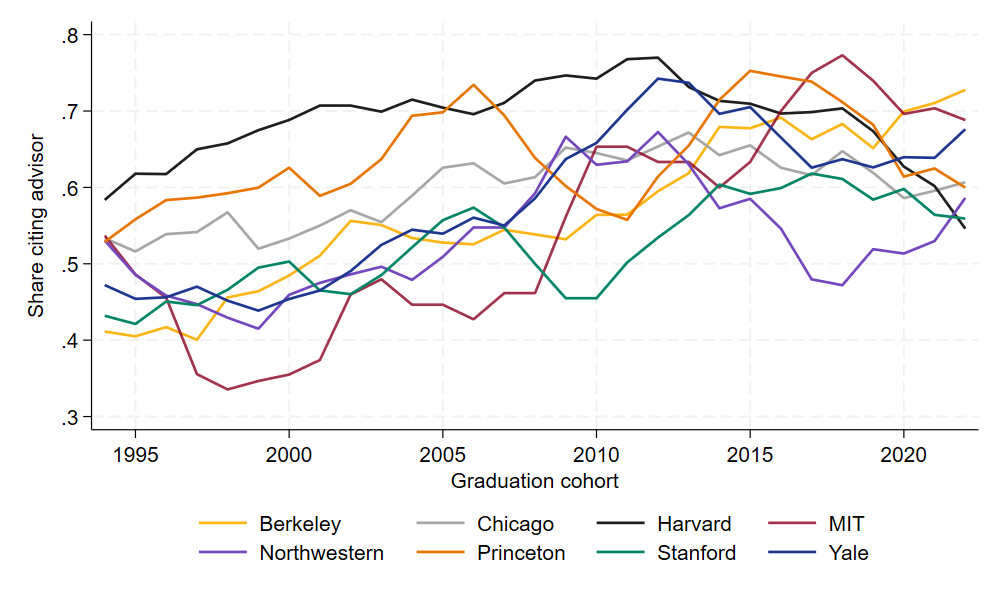}
	\label{fig: Students_citing_advisor}
	\begin{flushleft}
 \noindent\scriptsize{\emph{Notes}: This figure plots the share of PhD theses that cite an advisor's work, computed for a sample of roughly 6000 economics + related program graduates with thesis PDFs in the 1994-2023 cohorts. Data are smoothed using a five-year moving average. Citations are identified by titles in bibliographic material; see the data appendix for details.}
	\end{flushleft}
\end{figure}

The share of PhD theses citing an advisor's work has also trended upwards for most of the cohorts included in our samples. This trend is visible in the advisor citation rates plotted in Figure \ref{fig: Students_citing_advisor}, albeit with considerable cross-school variation and within-school variation over time. Citations to advisors have been highest at Harvard, peaking with over 70\% of early-2010s theses citing advisor work. Citations to advisors increased most at MIT, from an initial rate around 40\% to a rate similar to that of Harvard PhDs by 2015. Princeton PhDs' citation rates evolve much like those for Harvard PhDs, while Berkeley, Chicago, Northwestern, Stanford and Yale advisor-citation rates hover between 50-65\% by the end of our sample period. These patterns suggest growing research affinity between advisors and advisees in general, though with persistent differences in the degree of intellectual alignment across institutions.

Our measure of thesis citations to advisors is imperfect and relies in large part on algorithms that read thesis PDFs.\footnote{Thesis citations are not indexed by the Web of Science or EconLit. We therefore search thesis PDFs for all article titles of advisors' publication lists as detailed in the data appendix.} Advisor affiliation is also approximate in our data, relying on author affiliations attached to publications in EconLit. Still, the view that the resulting citation rates are informative is supported by Figure \ref{fig: transitions}. This figure tracks the number of students advised and student citations to advisors for advisors that change affiliations. For 364 fixed-super advisors (defined as an advisor whose advising load falls in the upper decile of the advisee distribution for any cohort), the figure counts advisees at the former affiliation and new affiliation separately. A transitioning advisor's advising load at their previous affiliation falls sharply at the time of a move, plateauing close to zero four years later. At the same time, the number of advisees advised by transitioning advisors at the new institution increases rapidly. A similar pattern appears in the number of graduates that cite transitioning advisors' work: their number decreases at a transitioning advisor's previous institution while it increases at the new one (considering all students at origin and destination schools, not just graduates advised by transitioning advisors).\footnote{Our work here is in the spirit of \cite{lerner2024wandering}, which traces patent citations to transitioning researchers' papers as they move across institutions.} These patterns suggest advisees citations capture a decline in intellectual influence at an advisor's old institution and rising influence at the new one.
 
When included in a regression model like equation \eqref{eq:studentmodel}, variables that indicate coauthoring with advisors and classmates before or in graduation years are unrelated to PhD students' immediate post-graduation publication success.  Estimated advisor-coauthoring effects on DI publications, reported in the first four columns of Table \ref{t: Coauthoring}, are small and not significantly different from zero. Classmate-coauthoring coefficient estimates, reported in columns 2 and 4 of the table, are on the margin or not significantly different from zero, though estimated much less precisely than the corresponding advisor-coauthoring effects.  

Graduates whose thesis cites an advisor, by contrast, are estimated to see $0.23$ more DI publications and $0.07$ more T6 publications, estimated gains that are significantly different from zero.  The Poisson estimates in columns 6 and 8 of Table \ref{t: Coauthoring} show that these gains amount to publication increases of 15\% and 22\%, respectively.  The strongest predictor of post-PhD publication success is a dummy variable indicating graduates with any pre-PhD publication; this is a control necessitated by the fact that our definition of coauthoring requires publication and precocious pre-degree publishers are more likely to publish after leaving the nest. 

As can be seen in the first three rows of Table \ref{t: Coauthoring}, estimated advisor effects generated by the extended version of equation \eqref{eq:studentmodel} with coauthoring and affinity variables are similar to those in the first three rows of Table 3, with significant estimates of coefficients on advisor research and past student success. Replacing continuous advisor attributes with dummies indicating \emph{fixed-duper} advisors generates the estimates in Table \ref{t: Coauthoring Fixed Duper}. In this model, fixed-duper dummies indicate advisors with a given attribute (e.g., number of advisees) ever coded as duper.
This specification is motivated by a view of exceptional advisor performance as a time-invariant attribute. The resulting estimates are broadly consistent with those in Tables \ref{t:multistudent} and \ref{t: Coauthoring}, suggesting that the most important advisor attributes are advisor research and past student success.  In contrast with the estimates in Table \ref{t: Coauthoring}, however, estimated fixed-duper advising load coefficients are larger--around $0.1$--and significantly different from zero for graduates' T6 publications. It's noteworthy, however, that regardless of specification, student success is hard to predict: for all models reported in Tables \ref{t:multistudent}-\ref{t: Coauthoring Fixed Duper}, $R^2$s are no bigger than .14 for DI publications and below .1 for T6.


\section{The Aggregate Research Production Function}\label{sec:aggregate} 

\subsection{Aggregation IV}

The strong relationship between advising features and student research success documented in Tables \ref{t:multistudent}-\ref{t: Coauthoring Fixed Duper} suggests factors like an advisor's research record and research affinity contribute to their advisees' success.  But estimates of this relationship may also reflect selection bias. Advisors whose students have done well in the past, for instance, may attract students who are most likely to succeed in the future. A school-level analysis mitigates this sort of bias by asking what happens to overall average success among graduates of, say, Princeton and Berkeley, when a prolific advisor moves from the former to the latter.

As before, let $c(i) \in \{1994 \dots 2017 \}$ encode graduation cohort for PhD $i$ and let $s(i)$ encode which of the eight schools $i$ attended. Suppose average potential research publication outcomes when $D_i=0$ can be described by the conditional expectation function (CEF):
\begin{equation}\label{eq:CEF}
E[R_{isc}(0)| c(i), s(i); W_i]=\tilde{\alpha}'W_i + \tilde{\gamma}_{c(i)} + \tilde{\delta}_{s(i)}.
\end{equation}
This says that, conditional on cohort and school, the CEF of potential research outcomes at a reference level of advising inputs is assumed to be an additive function of cohort and school effects, possibly with adjustment for $W_i$.  Suppose also that the causal effects of advising features vector $D_i$ (augmented to include coauthoring and research affinity variables) and advisor team size, $N_i$, on post-PhD research are constant and given by $\tilde{\tau}$ and $\tilde{\beta}$, respectively. These causal effects and the coefficients in \eqref{eq:CEF}, denoted by Greek letters with a $\tilde{}$, need not coincide with coefficients in an OLS estimand defined by a regression of $R_{isc}$ on advising features and controls, as in equation \eqref{eq:studentmodel}.  

Restriction \eqref{eq:CEF} and this constant effects assumption imply the following conditional moment restriction:
\begin{equation}\label{eq:IVID}
    E[R_{isc} - \tilde{\tau}'D_i  - \tilde{\beta} N_i - \tilde{\alpha}'W_i - \tilde{\gamma}_{c(i)} - \tilde{\delta}_{s(i)}| c(i), s(i); W_i]=0.
\end{equation} 
In other words, conditional on cohort and school effects, variation in the productivity of graduates by cohort and school is explained by variation in the mean of right-hand-side variables by cohort and school. Without individual controls, $W_i$ (in this case, dummies for female graduates, unclassified gender, economics department graduates and missing thesis PDFs), this moment restriction generates a grouped model that regresses average research productivity by cohort and school on average advisor research, the share of graduates coauthoring with advisors, and so on. With individual covariates, these regressors are adjusted for compositional changes due to a changing mix of gender, economics department, and PDF availability for the graduates in our data. 

\begin{figure}[h!]
	\centering
	\caption{The Effect of Advisor Affiliation Changes on Graduate Students Advised and Student Citations to Advisor Works}	
	\includegraphics[width=\linewidth]{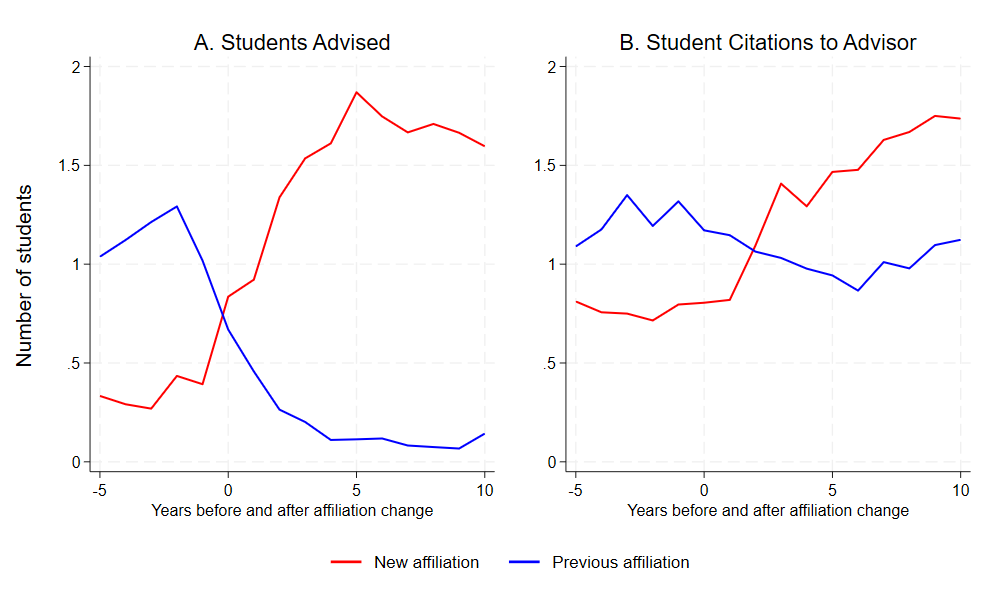}
	\label{fig: transitions}
	\begin{flushleft}
 \noindent\scriptsize{\emph{Notes}:  The left panel shows the number of graduates advised by advisors who change affiliations, with separate counts for cohorts graduating at the former and new affiliation before and after a transition year. The right panel plots the number of graduates citing a transitioning advisors' work at the previous and new affiliation. The analysis includes 364 fixed-super advisors (ranked by advisee count) with at least one affiliation change, as indicated by the affiliation variable on their EconLit publications. The sample is restricted to research-active advisors whose affiliations are observed in at least one-third of their research-active years (the years between their first and last publications). Advisors in this sample advised at least one student as of an affiliation change.}
	\end{flushleft}
\end{figure}

\begin{figure}[h!]
	\centering
	\caption{The Effect of Advisor Affiliation Changes on Average Cohort Publications}	
	\includegraphics[width=\linewidth]{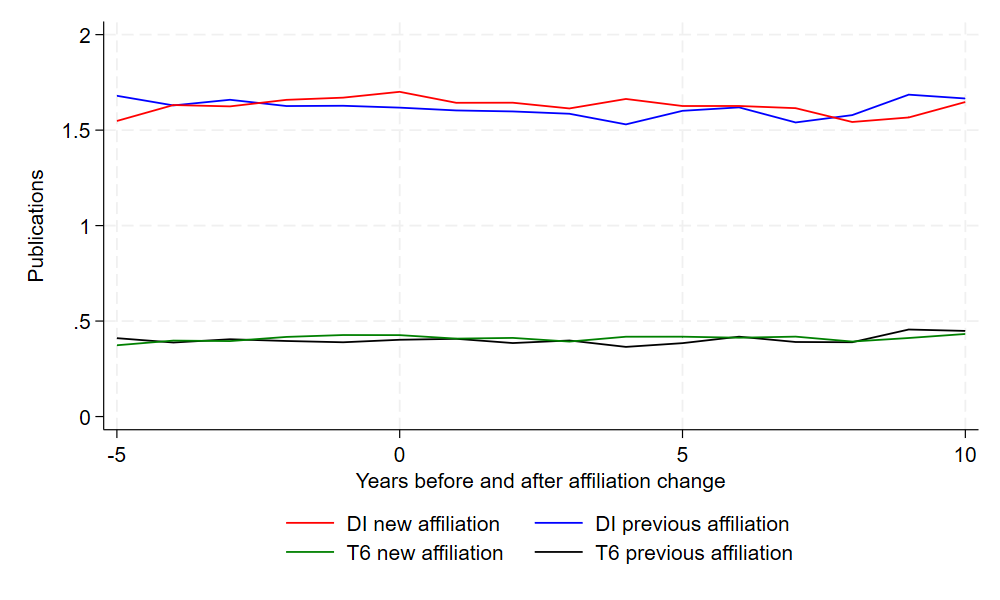}
	\label{fig: trans2}
	\begin{flushleft}
 \noindent\scriptsize{\emph{Notes}: This figure plots overall school-by-cohort averages of Deep Impact and Top 6 publication output in years $c+1$ to $c+6$ for graduates in cohort $c$. As in the previous figure, these are computed separately for schools losing or gaining a transitioning advisor, looking at cohorts completing degrees before and after a transition year. Transitioners here are 388 fixed-duper advisors (ranked by past student success) with at least one affiliation change, as indicated by the affiliation variable on their EconLit publications, limited to research-active advisors whose affiliations are observed in more than one-third of their research-active years, with at least one advisee graduating ahead of the affiliation change.  Publication statistics are for all graduates from affected schools.}
	\end{flushleft}
\end{figure}

Importantly, two-stage least squares (2SLS) estimates based on restriction \eqref{eq:IVID} are not confounded by \emph{within-school} sorting of students to advisors. Suppose, for instance, that within departments, the best students seek out advisors with strong research records, but advisor research success generates no payoff in terms of advisee research success.  In this scenario, estimates like those in Table \ref{t: Coauthoring} are likely to be positive: the fact that productive students seek out productive advisors engenders positive omitted variables bias. Yet, such within-school sorting leaves average student success by cohort and school unchanged, making the aggregate student research production function a better guide to causal advisor effects than equation \eqref{eq:studentmodel} estimated using data on individual graduates.     
Because the aggregate model controls for cohort and school effects, restriction \eqref{eq:IVID} identifies causal effects by exploiting cross-cohort \emph{changes} in average advising features within schools. These changes are due both to advisor transitions between schools and evolving within-department changes in incumbent advisor features. Elite departments compete to attract top scholars and prolific advisors, hoping (among other hiring goals) for an immediate boost in advising horsepower. As a suggestive exploration of the role played by advisor transitions, Figure \ref{fig: trans2} plots average research output for graduates at schools losing and receiving transitioning fixed-duper advisors as determined by their past student success. This figure is constructed in a manner similar to that used to construct Figure \ref{fig: transitions}. We opt here for fixed-duper rather than fixed-super advisor transitions because the former generate larger student-level estimates in Table \ref{t: Coauthoring Fixed Duper}. In contrast with Figure \ref{fig: transitions}, which shows advising load and citation changes consistent with transitioning advisors having an impact on affected schools, the pattern of student success traced in Figure \ref{fig: trans2} shows little evidence of gains in student research success at departments bolstered by the arrival of a fixed-duper advisor.   

In principle, restriction \eqref{eq:IVID} justifies a 2SLS estimator using a full set of 192 cohort-by-school dummies as instruments. Many of these instruments are weak, however, in the sense that they generate noisy first-stage estimates based on only a few students in each cohort for some schools. The resulting 2SLS estimates are therefore likely to be biased, and misleadingly similar to the corresponding OLS estimates. This motivates an IV strategy using 49 3-year-cohort-by-school dummies as instruments instead of the 192 dummies generated by single-year cohorts. The individual covariate vector, $W_i$, appears in both the first and second stages in this 2SLS setup.  Table \ref{t: 2SLS} reports 2SLS estimates in which all right-hand-side variables other than $W_i$ are instrumented as well as estimates from models in which only advisor research, advising load, and past student success are instrumented, treating other features like coauthoring and research affinity as covariates. 

Consistent with Figure \ref{fig: trans2}, 2SLS estimates using 49 dummy instruments provide less evidence that advising features matter than do the corresponding OLS estimates. 2SLS estimates with all features instrumented, reported in the first four columns of Table \ref{t: 2SLS}, generates statistically significant estimates only for the impact of advisor research.  Statistically significant 2SLS estimates of the impact of advisor research, on the order of $0.16$ for graduate DI publications and $0.047$ for graduate T6 publications, exceed the corresponding OLS estimates in columns 1-4 of Table \ref{t: Coauthoring} but are less precise. 2SLS estimates instrumenting only the three advisor attributes generate large, precisely-estimated pre-grad publication effects similar those in Table \ref{t: Coauthoring}. 

Estimates of coauthoring and advisor-citation effects in Table \ref{t: 2SLS} are so imprecise they should be seen as uninformative. On balance, this table does little to bolster claims for a strong relationship between advisor attributes and student success. Most notably, the coefficient on advising load is a reasonably-precisely estimated statistical zero and the coefficient on past student success is smaller than the effects seen in Table \ref{t: Coauthoring} and imprecisely estimated.  The strongest evidence for causal effects emerges for advisors' own research output, highlighting the critical role of advisor research success in shaping their advisees post-PhD research outcomes. Factors that seem to matter in Tables \ref{t: Coauthoring} and \ref{t: Coauthoring Fixed Duper} fail to generate consistent or precisely-estimated changes in average PhD student research success.

\subsection{Research Returns to Scale in Cohort Size} 

Motivated by the trend towards rising economics PhD program selectivity and falling cohort size at the super-elites, we conclude with an analysis of cohort size effects on student research success. In particular, we're interested in whether the scale of economics PhD research production function is constrained by decreasing returns.  Roughly speaking, cohort size can be thought of as graduate program class size. Perhaps reduced cohort size enhances students' post-graduation prospects by increasing resources and facilitating faculty-student mentoring. Larger cohorts, by contrast, may produce a critical mass of students in the classroom and more stimulating interactions between students. 

The relationship between cohort size and research productivity is quantified here using a regression of school-by-cohort aggregate graduate publication output up to six years after graduation on cohort size. This regression fits the following school-by-cohort CEF:
\begin{equation}\label{eq:scale}
   T_{sc}= \gamma_c + \delta_s + \kappa_1 n_{sc} + \kappa_2 n_{sc}^2 +\nu_sc,
\end{equation}
where $n_{sc}$ is the size of cohort $c$ at school $s$ and the dependent variable, $T_{sc}$, sums DI or T6 publications in years $c+1$ to $c+6$ by cohort-$c$ graduates from programs at $s$. Decreasing returns to scale in graduate student research are evinced by a negative estimate of $\kappa_2$, the coefficient on $n_{sc}^2$.

Estimates of of a linear version of equation \ref{eq:scale} using the full sample, reported in columns 1-2 and 5-6 of Panel A in Table \ref{t: CRTS}, indicate that DI publications increase by $1.4-1.6$ per student, while T6 publications increase by around $0.4$. These estimates are close to the mean post-publication statistics in Appendix Table \ref{t:appendixstats}. Estimates are similar when computed with and without school effects. Results using the sample of economics department graduates only, reported in Panel B, are also similar to those for the full sample.\footnote{Cohort size in this case is the number of identifiable economics department graduates.}

Estimates of $\kappa_2$ in models that include $n_{sc}^2$ suggest the graduate research production function is remarkably linear. Models with and without school effects, for the full sample and for economics department graduates only, generate estimates of $\kappa_2$ that are small and not significantly different from zero.  Inclusion of a quadratic term makes estimates of linear terms less precise, and nonlinear models for T6 publications are unstable and sensitive to the inclusion of school effects. Although the estimates for T6 output are sensitive to the inclusion of school effects, the results are reasonably consistent across specifications and samples. 

The estimated marginal effect of cohort size in equation \eqref{eq:scale} is $\hat{\kappa}_1+2\hat{\kappa}_2 n_{sc}$. The estimated change in returns to scale as a function of cohort size is therefore $2\hat{\kappa}_2$. Estimates of $\kappa_2$ for DI publications in the full sample imply a modest effect-change gradient of $-.0132 (2\times-0.0066)$ when estimated without school effects. This reduces to $-.003$ when school effects are added. The implied change in returns to scale as enrollment grows are therefore very small.   

On balance, the estimates in Table \ref{t: CRTS} suggest that departments looking to increase social impact through academic economics research can do so cheaply and quickly by admitting more students. This possible free lunch likely partly reflects the fact that advising is so concentrated. The schools in our sample appear to have plenty of advising slack.

\section{Summary and Conclusions: Selection, Training, and Luck on the Path to PhD}\label{sec:conclusion}

Top economics departments attract exceptional students and invest substantial resources in training these students for successful research careers. The most important graduate education resource is the time and attention of PhD advisors. With the help of uniquely rich and comprehensive data linking economics PhD students with their advisors, our analysis quantifies the relationship between key features of advisors and the advising relationship and the post-degree research success of economics graduate students.

Key descriptive facts emerging from our analysis include the high concentration of advising among a minority of advisors and the limited research success seen by the median graduate. It's also noteworthy that, even as elite programs have grown costlier and more selective, graduate research success has remained reasonably flat across cohorts. Other descriptive findings include the fact that research performance differs little between identifiable economics department graduates and their peers from economics-related programs. We've also shown that, after a brief warm-up period characterized by gender parity in research output, female graduates publish fewer papers than do male graduates. But the gender gap in research output, which is unrelated to advisor gender or advisor-advisee gender matching, may now be closing.

Which factors increase the likelihood of post-PhD research success? Multivariate models that predict graduate research success as a function of advisor and thesis characteristics yield robust positive effects of advisor research and the advising team's past student success. Surprisingly, however, a relationship between an advising team's advising load and student research success emerges only for dummy variables indicating the most prolific advisors. We've also seen that while precocious pre-PhD student publishers publish more papers post-PhD, research output appears to be unrelated to coauthoring with advisors or classmates. On the other hand, PhDs who cite an advisor in their thesis--a measure of student-advisor research topic affinity--tend to see greater research success post-PhD.  

The possibility of advisor-advisee sorting and the attendant selection bias motivates 2SLS analysis using dummies for cohort-by-school to instrument advising features.  2SLS estimation uncovers only weak evidence of causal effects of advising features on PhD student research success. In particular, at the level of cohorts and schools, only average advisor research predicts research success for the average graduate.  2SLS estimates of effects of past student success and research affinity are uniformly positive but not significantly different from zero. 

We've also explored the nature of returns to scale in graduate economics education. Our analysis suggests that  aggregate student research output scales roughly linearly with graduate economics enrollment. This finding challenges conventional wisdom regarding the importance of small, highly selective cohorts in economics graduate education. On the margin, graduates in larger cohorts publish about as well as graduates in smaller cohorts.  

A broader lesson suggested by our findings is that research success is hard for elite schools to engineer or even predict. In this, academic economics is like professional sports: coaches at all levels struggle to identify and nurture talent; uncertainty regarding outcomes and performance is pervasive. Even among American Division I college basketball players, who are well-trained and necessarily play a very good game, few ultimately play for a living.  As in high-level sports, economics PhD students and their advisors alike should benefit from a clear-eyed view of the winding and uncertain path to research success.

\pagebreak
\section*{Exhibits}
\subsection*{Figures}

\clearpage

\subsection*{Tables}

\begin{table}[h]
    \centering 
    \caption{Cohort Size and Activity Rates}
    \label{t:gradcounts}
    \begin{tabular}{lcccccc}
\toprule
& \multicolumn{3}{c}{Economics} & \multicolumn{3}{c}{Related} \\
\cmidrule(lr){2-4} \cmidrule(lr){5-7}
& Total & Cohort size & Active & Total & Cohort size & Active \\
& (1) & (2) & (3) & (4) & (5) & (6)  \\ 
\toprule
\\ \multicolumn{7}{l}{Panel A. 1989-2023 Graduates} \\\\ [-1ex]
Berkeley&1119&32.0&&158&4.5&\\
Chicago&897&25.6&&155&4.4&\\
Harvard&1139&32.5&&169&4.8&\\
MIT&865&24.7&&217&6.2&\\
Northwestern&675&19.3&&75&2.1&\\
Princeton&650&18.6&&39&1.1&\\
Stanford&759&21.7&&325&9.3&\\
Yale&660&18.9&&101&2.9&\\
All&6764&24.2&&1239&4.4&\\
\\ \multicolumn{7}{l}{Panel B. 1994-2017 Graduates} \\\\ [-1ex]
Berkeley&854&35.6&.46&103&4.3&.36\\
Chicago&681&28.4&.38&105&4.4&.53\\
Harvard&791&33.0&.54&91&3.8&.42\\
MIT&631&26.3&.58&150&6.3&.31\\
Northwestern&485&20.2&.54&54&2.3&.50\\
Princeton&467&19.5&.57&27&1.1&.44\\
Stanford&530&22.1&.48&216&9.0&.46\\
Yale&454&18.9&.52&52&2.2&.29\\
All&4893&25.5&.50&798&4.2&.41\\
\bottomrule
\end{tabular}

    \begin{flushleft}
		\noindent\scriptsize{\emph{Notes}: Columns 1 and 4 show the number of PhDs awarded for economics subject matter from 1989-2023 and 1994-2017 for graduates of economics and related programs. Column 2 shows the average economics program cohort size; column 5 shows the average number of related-program graduates. Columns 3 and 6 in Panel B show the share of graduates who are active post-PhD, meaning they have at least one Deep Impact publication in the six years after graduation (years $c+1$ to $c+6$ in cohort $c$).} 
	\end{flushleft}
\end{table}

\begin{table}[H]
    \centering
      \caption{Students of Research-Active, Prolific Advisors Do Better}
   \begin{tabular}{ll}
    \resizebox{\columnwidth}{!}{
        \label{t:studentregs}
\def\sym#1{\ifmmode^{#1}\else\(^{#1}\)\fi} \begin{tabular}{lcccccccc} \hline \hline
& \multicolumn{4}{c}{Levels} & \multicolumn{4}{c}{Poisson} \\
\cmidrule(lr){2-5} \cmidrule(lr){6-9}
&\multicolumn{2}{c}{DI}&\multicolumn{2}{c}{T6}&\multicolumn{2}{c}{DI}&\multicolumn{2}{c}{T6} \\
\cmidrule(lr){2-3} \cmidrule(lr){4-5} \cmidrule(lr){6-7} \cmidrule(lr){8-9}
&\multicolumn{1}{c}{(1)}&\multicolumn{1}{c}{(2)}&\multicolumn{1}{c}{(3)}&\multicolumn{1}{c}{(4)}&\multicolumn{1}{c}{(5)}&\multicolumn{1}{c}{(6)}&\multicolumn{1}{c}{(7)}&\multicolumn{1}{c}{(8)}\\ 
\hline \\ A. Research Productivity \\
Advisor Research&     0.12\sym{***}&                  &    0.046\sym{***}&                  &    0.068\sym{***}&                  &    0.087\sym{***}&                  \\
                &  (0.010)         &                  & (0.0046)         &                  & (0.0056)         &                  & (0.0090)         &                  \\
[1em]
Super Advisor Research&                  &     0.70\sym{***}&                  &     0.24\sym{***}&                  &     0.47\sym{***}&                  &     0.67\sym{***}\\
                &                  &  (0.070)         &                  &  (0.026)         &                  &  (0.048)         &                  &  (0.081)         \\
B. Advising Load \\
Advising Load   &    0.033\sym{***}&                  &    0.015\sym{***}&                  &    0.020\sym{***}&                  &    0.032\sym{***}&                  \\
                & (0.0063)         &                  & (0.0027)         &                  & (0.0036)         &                  & (0.0055)         &                  \\
[1em]
Super Advising Load&                  &     0.47\sym{***}&                  &     0.19\sym{***}&                  &     0.34\sym{***}&                  &     0.70\sym{***}\\
                &                  &  (0.077)         &                  &  (0.031)         &                  &  (0.060)         &                  &   (0.13)         \\
C. Successful Students \\
Past Student Success&     0.43\sym{***}&                  &     0.15\sym{***}&                  &     0.25\sym{***}&                  &     0.35\sym{***}&                  \\
                &  (0.038)         &                  &  (0.016)         &                  &  (0.021)         &                  &  (0.034)         &                  \\
[1em]
Super Past Student Success&                  &     0.57\sym{***}&                  &     0.20\sym{***}&                  &     0.36\sym{***}&                  &     0.49\sym{***}\\
                &                  &  (0.068)         &                  &  (0.030)         &                  &  (0.042)         &                  &  (0.074)         \\
\hline
Dep. var. mean  &     1.56         &                  &     0.39         &                  &     1.56         &                  &     0.39         &                  \\
\hline\hline  \end{tabular}
}
    \end{tabular}
    \begin{flushleft}
		\noindent\scriptsize{\emph{Notes}: The dependent variable in columns 1, 2, 5, and 6 is student Deep Impact publications in the six years after graduation ($c+1$ to $c+6$ for cohort $c$). The dependent variable in columns 3, 4, 7, and 8 sums top-6 publications. \emph{Research productivity} regressors average a cohort-$c$ graduate's advisors' Deep Impact publications in the five years before $c$. \emph{Advising load} regressors average a graduate's advisors' advisee counts in the five years before $c$.  For a given graduate, the \emph{successful student} regressor averages the number of Deep Impact publications by a graduate's advisors' past advisees as described by equation \ref{eq:definePS} in the text. ``Super'' regressors are dummies indicating whether at least one of a student's advisors was among the 10\% most prolific advisors relative to all other advisors who advised students graduating in the five years preceding the student's graduation year. Models for each panel are run separately.  All models control for school and cohort effects, graduates' advisor team size, graduate gender, unclassified gender, missing thesis PDF, and an economics department dummy. Estimates are for the economics + related sample of 5676 students that graduated between 1994 and 2017. Standard errors are clustered on school-by-cohort (192 clusters for 8 schools $\times$ 24 cohorts).} 
	\end{flushleft}
\end{table}

\begin{table}[H]
    \centering
    \caption{Multivariate Model of Advisor Effects}
    \label{t:multistudent}
    \resizebox{\columnwidth}{!}{
   \begin{tabular}{ll}    \def\sym#1{\ifmmode^{#1}\else\(^{#1}\)\fi} \begin{tabular}{lcccccccc} \hline \hline
& \multicolumn{4}{c}{Levels} & \multicolumn{4}{c}{Poisson} \\
\cmidrule(lr){2-5} \cmidrule(lr){6-9} 
&\multicolumn{2}{c}{DI}&\multicolumn{2}{c}{T6}&\multicolumn{2}{c}{DI}&\multicolumn{2}{c}{T6} \\
\cmidrule(lr){2-3} \cmidrule(lr){4-5} \cmidrule(lr){6-7} \cmidrule(lr){8-9}
&\multicolumn{1}{c}{(1)}&\multicolumn{1}{c}{(2)}&\multicolumn{1}{c}{(3)}&\multicolumn{1}{c}{(4)}&\multicolumn{1}{c}{(5)}&\multicolumn{1}{c}{(6)}&\multicolumn{1}{c}{(7)}&\multicolumn{1}{c}{(8)}\\ 
\hline
Advisor Research&    0.100\sym{***}&                  &    0.036\sym{***}&                  &    0.055\sym{***}&                  &    0.066\sym{***}&                  \\
                &  (0.012)         &                  & (0.0052)         &                  & (0.0066)         &                  & (0.0099)         &                  \\
[1em]
Advising Load   &  -0.0083         &                  &   0.0010         &                  &  -0.0026         &                  &   0.0054         &                  \\
                & (0.0075)         &                  & (0.0031)         &                  & (0.0049)         &                  & (0.0069)         &                  \\
[1em]
Past Student Success&     0.30\sym{***}&                  &     0.10\sym{***}&                  &     0.19\sym{***}&                  &     0.27\sym{***}&                  \\
                &  (0.041)         &                  &  (0.017)         &                  &  (0.022)         &                  &  (0.035)         &                  \\
[1em]
Super Advisor Research&                  &     0.46\sym{***}&                  &     0.12\sym{***}&                  &     0.33\sym{***}&                  &     0.40\sym{***}\\
                &                  &   (0.10)         &                  &  (0.039)         &                  &  (0.066)         &                  &   (0.11)         \\
[1em]
Super Advising Load&                  &     0.15         &                  &    0.068\sym{*}  &                  &     0.13\sym{**} &                  &     0.36\sym{**} \\
                &                  &  (0.088)         &                  &  (0.035)         &                  &  (0.067)         &                  &   (0.14)         \\
[1em]
Super Past Student Success&                  &     0.28\sym{***}&                  &    0.093\sym{***}&                  &     0.19\sym{***}&                  &     0.25\sym{***}\\
                &                  &  (0.075)         &                  &  (0.033)         &                  &  (0.046)         &                  &  (0.082)         \\
[1em]
Duper Advisor Research&                  &     0.58\sym{***}&                  &     0.20\sym{***}&                  &     0.38\sym{***}&                  &     0.52\sym{***}\\
                &                  &  (0.079)         &                  &  (0.030)         &                  &  (0.053)         &                  &  (0.085)         \\
[1em]
Duper Advising Load&                  &     0.29\sym{***}&                  &     0.14\sym{***}&                  &     0.22\sym{***}&                  &     0.52\sym{***}\\
                &                  &  (0.082)         &                  &  (0.034)         &                  &  (0.063)         &                  &   (0.14)         \\
[1em]
Duper Past Student Success&                  &     0.61\sym{***}&                  &     0.21\sym{***}&                  &     0.35\sym{***}&                  &     0.47\sym{***}\\
                &                  &   (0.10)         &                  &  (0.042)         &                  &  (0.055)         &                  &  (0.090)         \\
\hline
Dep. var. mean  &     1.56         &                  &     0.39         &                  &     1.56         &                  &     0.39         &                  \\
R2              &    0.089         &    0.076         &    0.078         &    0.068         &                  &                  &                  &                  \\
\hline\hline  \end{tabular}

    \end{tabular}}
    \begin{flushleft}
		\noindent\scriptsize{\emph{Notes}: Dependent variables and sample are as described in the note to Table \ref{t:studentregs}. Each column reports estimates from a single regression model with multiple advisor attributes on the right-hand side.  Super and duper dummies indicate graduates with at least one advisor with characteristics in percentiles 6-10 (super) and  1-5 (duper). Standard errors are clustered on school-by-cohort (192 clusters).  Models control for school and cohort effects, advisor team size, graduate gender,  unclassified gender,  missing thesis PDF, and an economics department dummy.} 
	\end{flushleft}
\end{table}

\begin{table}[H]
    \centering
    \caption{Effects of Coauthoring and Research Affinity, Continuous Advisor Attributes}
    \label{t: Coauthoring}
    \resizebox{\columnwidth}{!}{
   \begin{tabular}{ll}
        \def\sym#1{\ifmmode^{#1}\else\(^{#1}\)\fi} \begin{tabular}{lcccccccc} \hline \hline
& \multicolumn{4}{c}{Levels} & \multicolumn{4}{c}{Poisson} \\
\cmidrule(lr){2-5} \cmidrule(lr){6-9}
&\multicolumn{2}{c}{DI}&\multicolumn{2}{c}{T6}&\multicolumn{2}{c}{DI}&\multicolumn{2}{c}{T6} \\
\cmidrule(lr){2-3} \cmidrule(lr){4-5} \cmidrule(lr){6-7} \cmidrule(lr){8-9}
&\multicolumn{1}{c}{(1)}&\multicolumn{1}{c}{(2)}&\multicolumn{1}{c}{(3)}&\multicolumn{1}{c}{(4)}&\multicolumn{1}{c}{(5)}&\multicolumn{1}{c}{(6)}&\multicolumn{1}{c}{(7)}&\multicolumn{1}{c}{(8)}\\ 
\hline
Advisor Research&    0.095\sym{***}&    0.089\sym{***}&    0.034\sym{***}&    0.032\sym{***}&    0.052\sym{***}&    0.049\sym{***}&    0.061\sym{***}&    0.057\sym{***}\\
                &  (0.012)         &  (0.012)         & (0.0051)         & (0.0050)         & (0.0065)         & (0.0064)         & (0.0098)         & (0.0096)         \\
[1em]
Advising Load   &  -0.0052         &  -0.0045         &   0.0022         &   0.0024         & -0.00072         & -0.00018         &   0.0087         &   0.0094         \\
                & (0.0074)         & (0.0073)         & (0.0030)         & (0.0030)         & (0.0048)         & (0.0047)         & (0.0067)         & (0.0066)         \\
[1em]
Past Student Success&     0.29\sym{***}&     0.28\sym{***}&    0.097\sym{***}&    0.094\sym{***}&     0.18\sym{***}&     0.18\sym{***}&     0.26\sym{***}&     0.26\sym{***}\\
                &  (0.040)         &  (0.040)         &  (0.017)         &  (0.017)         &  (0.021)         &  (0.021)         &  (0.034)         &  (0.034)         \\
[1em]
Coauthored with Advisor Pre-grad&    -0.12         &    -0.15         &     0.23         &     0.22         &   -0.061         &   -0.080         &     0.20         &     0.17         \\
                &   (0.24)         &   (0.24)         &   (0.15)         &   (0.15)         &  (0.078)         &  (0.079)         &   (0.15)         &   (0.15)         \\
[1em]
Coauthored with Classmate Pre-grad&                  &     0.73\sym{*}  &                  &     0.28         &                  &     0.18         &                  &     0.22         \\
                &                  &   (0.44)         &                  &   (0.26)         &                  &   (0.12)         &                  &   (0.24)         \\
[1em]
Any Publication Pre-grad&     1.73\sym{***}&     1.64\sym{***}&     0.46\sym{***}&     0.43\sym{***}&     0.78\sym{***}&     0.75\sym{***}&     0.81\sym{***}&     0.78\sym{***}\\
                &   (0.15)         &   (0.16)         &  (0.068)         &  (0.071)         &  (0.051)         &  (0.055)         &  (0.085)         &  (0.089)         \\
[1em]
Cites Advisor in Thesis&                  &     0.23\sym{***}&                  &    0.065\sym{**} &                  &     0.15\sym{***}&                  &     0.22\sym{***}\\
                &                  &  (0.068)         &                  &  (0.027)         &                  &  (0.046)         &                  &  (0.078)         \\
\hline
Dep. var. mean  &     1.56         &                  &     0.39         &                  &     1.56         &                  &     0.39         &                  \\
R2              &     0.14         &     0.14         &     0.11         &     0.11         &                  &                  &                  &                  \\
\hline\hline  \end{tabular}

    \end{tabular}}
    \begin{flushleft}
        \noindent\scriptsize{\emph{Notes}: Dependent variables and sample are as described in the note to Table \ref{t:studentregs}. Models control for school and cohort effects, graduates' advisor team size, graduate gender,  unclassified gender, missing thesis PDF, and an economics department dummy. Standard errors are clustered on school-by-cohort (192 clusters). Of the 5676 economics+related program graduates in the 1994-2017 cohorts, 575 published by the end of graduation year. Of those, 193 coauthored with an advisor and 61 with a classmate. \textit{Cites Advisor in Thesis} is a dummy variable indicating PhD theses that cite an advisor's work.} 
    \end{flushleft}
\end{table}

\begin{table}[H]
    \centering
    \caption{Effects of Coauthoring and Research Affinity, Fixed-Duper Advisors}
    \label{t: Coauthoring Fixed Duper}
    \resizebox{\columnwidth}{!}{
   \begin{tabular}{ll}
        \def\sym#1{\ifmmode^{#1}\else\(^{#1}\)\fi} \begin{tabular}{lcccccccc} \hline \hline
& \multicolumn{4}{c}{Levels} & \multicolumn{4}{c}{Poisson} \\
\cmidrule(lr){2-5} \cmidrule(lr){6-9}
&\multicolumn{2}{c}{DI}&\multicolumn{2}{c}{T6}&\multicolumn{2}{c}{DI}&\multicolumn{2}{c}{T6} \\
\cmidrule(lr){2-3} \cmidrule(lr){4-5} \cmidrule(lr){6-7} \cmidrule(lr){8-9}
&\multicolumn{1}{c}{(1)}&\multicolumn{1}{c}{(2)}&\multicolumn{1}{c}{(3)}&\multicolumn{1}{c}{(4)}&\multicolumn{1}{c}{(5)}&\multicolumn{1}{c}{(6)}&\multicolumn{1}{c}{(7)}&\multicolumn{1}{c}{(8)}\\ 
\hline
Fixed-Duper Advisor Research&     0.49\sym{***}&     0.45\sym{***}&     0.16\sym{***}&     0.14\sym{***}&     0.35\sym{***}&     0.33\sym{***}&     0.53\sym{***}&     0.50\sym{***}\\
                &  (0.075)         &  (0.075)         &  (0.027)         &  (0.027)         &  (0.059)         &  (0.059)         &   (0.11)         &   (0.11)         \\
[1em]
Fixed-Duper Advising Load&    0.034         &    0.031         &    0.097\sym{***}&    0.096\sym{***}&    0.054         &    0.055         &     0.48\sym{***}&     0.49\sym{***}\\
                &  (0.083)         &  (0.083)         &  (0.028)         &  (0.028)         &  (0.068)         &  (0.069)         &   (0.14)         &   (0.14)         \\
[1em]
Fixed-Duper Past Student Success&     0.86\sym{***}&     0.83\sym{***}&     0.19\sym{***}&     0.18\sym{***}&     0.84\sym{***}&     0.82\sym{***}&     0.94\sym{***}&     0.91\sym{***}\\
                &  (0.067)         &  (0.066)         &  (0.022)         &  (0.021)         &  (0.072)         &  (0.072)         &   (0.12)         &   (0.12)         \\
[1em]
Coauthored with Advisor Pre-grad&    -0.16         &    -0.20         &     0.23         &     0.21         &   -0.090         &    -0.11         &     0.18         &     0.15         \\
                &   (0.24)         &   (0.24)         &   (0.16)         &   (0.16)         &  (0.076)         &  (0.076)         &   (0.15)         &   (0.15)         \\
[1em]
Coauthored with Classmate Pre-grad&                  &     0.76\sym{*}  &                  &     0.30         &                  &     0.20\sym{*}  &                  &     0.25         \\
                &                  &   (0.44)         &                  &   (0.26)         &                  &   (0.12)         &                  &   (0.23)         \\
[1em]
Any Publication Pre-grad&     1.75\sym{***}&     1.66\sym{***}&     0.47\sym{***}&     0.44\sym{***}&     0.80\sym{***}&     0.77\sym{***}&     0.84\sym{***}&     0.80\sym{***}\\
                &   (0.15)         &   (0.15)         &  (0.068)         &  (0.071)         &  (0.050)         &  (0.052)         &  (0.081)         &  (0.085)         \\
[1em]
Cites Advisor in Thesis&                  &     0.27\sym{***}&                  &    0.093\sym{***}&                  &     0.17\sym{***}&                  &     0.26\sym{***}\\
                &                  &  (0.066)         &                  &  (0.027)         &                  &  (0.044)         &                  &  (0.077)         \\
\hline
Dep. var. mean  &     1.56         &                  &     0.39         &                  &     1.56         &                  &     0.39         &                  \\
R2              &     0.13         &     0.14         &    0.092         &    0.095         &                  &                  &                  &                  \\
\hline\hline  \end{tabular}

    \end{tabular}}
    \begin{flushleft}
        \noindent\scriptsize{\emph{Notes}: Dependent variables and sample are as described in the note to Table \ref{t:studentregs}. Models control for school and cohort effects, graduates' advisor team size, graduate gender,  unclassified gender, missing thesis PDF, and an economics department dummy. Fixed-duper dummies indicate advisors ever coded duper for a given advisor attribute. Within advisors, fixed-duper dummies are cohort-invariant.} 
    \end{flushleft}
\end{table}

\begin{table}[H]
    \centering
    \caption{2SLS Estimates Using 49 School $\times$ Cohort Dummies as Instruments}
    \label{t: 2SLS}
    \resizebox{\columnwidth}{!}{
   \begin{tabular}{ll}
        \def\sym#1{\ifmmode^{#1}\else\(^{#1}\)\fi} \begin{tabular}{lcccccccc} \hline \hline
& \multicolumn{4}{c}{All features instrumented} & \multicolumn{4}{c}{First 3 instrumented} \\
\cmidrule(lr){2-5} \cmidrule(lr){6-9} 
&\multicolumn{2}{c}{DI}&\multicolumn{2}{c}{T6}&\multicolumn{2}{c}{DI}&\multicolumn{2}{c}{T6} \\
\cmidrule(lr){2-3} \cmidrule(lr){4-5} \cmidrule(lr){6-7} \cmidrule(lr){8-9}
&\multicolumn{1}{c}{(1)}&\multicolumn{1}{c}{(2)}&\multicolumn{1}{c}{(3)}&\multicolumn{1}{c}{(4)}&\multicolumn{1}{c}{(5)}&\multicolumn{1}{c}{(6)}&\multicolumn{1}{c}{(7)}&\multicolumn{1}{c}{(8)}\\ 
\hline
Advisor Research&     0.16\sym{***}&     0.16\sym{***}&    0.045\sym{*}  &    0.047\sym{*}  &     0.17\sym{***}&     0.16\sym{***}&    0.052\sym{**} &    0.051\sym{**} \\
                &  (0.050)         &  (0.048)         &  (0.024)         &  (0.024)         &  (0.048)         &  (0.047)         &  (0.023)         &  (0.022)         \\
[1em]
Advising Load   &   -0.020         &   -0.020         &   -0.010         &   -0.010         &   -0.029         &   -0.029         &   -0.016         &   -0.016         \\
                &  (0.025)         &  (0.023)         &  (0.012)         &  (0.012)         &  (0.024)         &  (0.023)         &  (0.011)         &  (0.011)         \\
[1em]
Past Student Success&     0.12         &    0.049         &    0.086         &    0.073         &     0.14         &     0.11         &     0.11         &    0.097         \\
                &   (0.19)         &   (0.20)         &  (0.072)         &  (0.076)         &   (0.17)         &   (0.18)         &  (0.068)         &  (0.068)         \\
[1em]
Coauthored with Advisor Pre-grad&     0.15         &    -0.23         &     1.03         &     0.95         &    -0.14         &    -0.16         &     0.22         &     0.21         \\
                &   (2.04)         &   (2.06)         &   (0.89)         &   (0.90)         &   (0.24)         &   (0.24)         &   (0.15)         &   (0.15)         \\
[1em]
Coauthored with Classmate Pre-grad&                  &     0.55         &                  &     0.47         &                  &     0.71\sym{*}  &                  &     0.26         \\
                &                  &   (3.04)         &                  &   (1.43)         &                  &   (0.43)         &                  &   (0.26)         \\
[1em]
Any Publication Pre-grad&     1.10         &     0.98         &     0.35         &     0.29         &     1.72\sym{***}&     1.64\sym{***}&     0.46\sym{***}&     0.43\sym{***}\\
                &   (1.18)         &   (1.24)         &   (0.49)         &   (0.52)         &   (0.15)         &   (0.16)         &  (0.068)         &  (0.071)         \\
[1em]
Cites Advisor in Thesis&                  &     0.96         &                  &     0.13         &                  &     0.18\sym{*}  &                  &    0.051         \\
                &                  &   (0.64)         &                  &   (0.27)         &                  &   (0.10)         &                  &  (0.046)         \\
\hline
Dep. var. mean  &     1.56         &                  &     0.39         &                  &     1.56         &                  &     0.39         &                  \\
\hline\hline  \end{tabular}

    \end{tabular}}
    \begin{flushleft}
        \noindent\scriptsize{\emph{Notes}: The first 4 columns report 2SLS estimates computed using 49 dummies for schools $\times$ 3-year cohorts as instruments, with all listed variables plus advisor team size instrumented. Estimates in columns 5-8 are from models instrumenting the first three advisor attributes only. Models control for school and cohort effects, graduates' advisor team size (instrumented in the first 4 columns), graduate gender,  unclassified gender, missing thesis PDF, and an economics department dummy. Standard errors are clustered on school-by-cohort.
        } 
    \end{flushleft}
\end{table}

\begin{table}[H]
    \centering
    \caption{Effects of Cohort Size on Aggregate Research Productivity}
    \label{t: CRTS}
    \resizebox{\columnwidth}{!}{
    \begin{tabular}{ll}\def\sym#1{\ifmmode^{#1}\else\(^{#1}\)\fi} \begin{tabular}{l*{8}{l}} \hline\hline
& \multicolumn{4}{c}{Deep Impact} & \multicolumn{4}{c}{Top 6} \\
\cmidrule(lr){2-5} \cmidrule(lr){6-9}
                &\multicolumn{1}{c}{(1)}         &\multicolumn{1}{c}{(2)}         &\multicolumn{1}{c}{(3)}         &\multicolumn{1}{c}{(4)}         &\multicolumn{1}{c}{(5)}         &\multicolumn{1}{c}{(6)}         &\multicolumn{1}{c}{(7)}         &\multicolumn{1}{c}{(8)}         \\
\hline \\ \multicolumn{8}{l}{{Panel A. Economics + Related Graduates }} \\\\[-1ex]
Cohort Size     &     1.39\sym{***}&     1.57\sym{***}&     1.79\sym{***}&     1.67\sym{***}&     0.46\sym{***}&     0.36\sym{***}&     0.87\sym{***}&     0.31         \\
                &   (0.13)         &   (0.18)         &   (0.65)         &   (0.59)         &  (0.064)         &  (0.081)         &   (0.31)         &   (0.25)         \\
[1em]
Cohort Size Squared&                  &                  &  -0.0066         &  -0.0015         &                  &                  &  -0.0068         &  0.00087         \\
                &                  &                  &  (0.011)         & (0.0097)         &                  &                  & (0.0052)         & (0.0042)         \\
\multicolumn{8}{l}{{Panel B. Economics Graduates}} \\\\[-2ex]
Cohort Size     &     1.46\sym{***}&     1.65\sym{***}&     2.02\sym{***}&     1.68\sym{**} &     0.48\sym{***}&     0.40\sym{***}&     0.87\sym{**} &     0.16         \\
                &   (0.16)         &   (0.21)         &   (0.76)         &   (0.68)         &  (0.078)         &  (0.095)         &   (0.36)         &   (0.29)         \\
[1em]
Cohort Size Squared&                  &                  &   -0.010         & -0.00039         &                  &                  &  -0.0069         &   0.0043         \\
                &                  &                  &  (0.014)         &  (0.013)         &                  &                  & (0.0067)         & (0.0056)         \\
\hline
School effects  &       No         &      Yes         &       No         &      Yes         &       No         &      Yes         &       No         &      Yes         \\
\hline\hline  \end{tabular} 

    \end{tabular}
    }
    \begin{flushleft}
		\noindent\scriptsize{\emph{Notes}: This table reports estimates from regressions of school-by-cohort average publication output in $c+1$ to $c+6$ on the size of cohort $c$. Dependent variables sum either Deep Impact or Top 6 publications for cohorts graduating 1994-2017. The estimation sample consists of 192 school-by-year groups. Panel A reports estimates for models averaging the sample of economics+related graduates. Panel B reports estimates for averages computed using identifiable economics department graduates only. All specifications include cohort effects. Models reported in even-numbered columns include school effects. Robust standard errors appear in parentheses.} 
	\end{flushleft}
\end{table}

\clearpage
\section*{Appendix: Additional Exhibits}
\setcounter{table}{0}
\renewcommand{\thetable}{A\arabic{table}}
\setcounter{figure}{0}
\renewcommand{\thefigure}{A\arabic{figure}}

\begin{figure}[h!]
	\centering
 	\caption{Advisor and Advisee Publication Shares}
    \includegraphics[width=\linewidth, keepaspectratio]{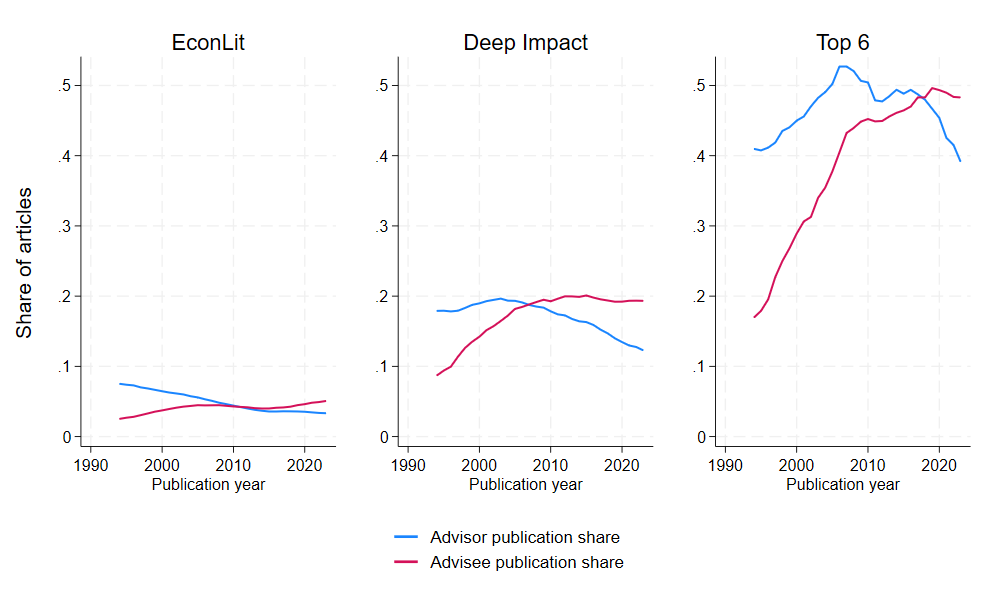}
  	\label{fig: Intro_2024_v2}
	\begin{flushleft}
		\noindent\scriptsize{\emph{Notes}: This figure plots a five-year moving average of the yearly share of publications authored or coauthored by economics and related program graduates earning degrees at one of the eight institutions in our sample, along with the share of publications authored by their advisors. Papers coauthored by advisors and advisees contributed to the shares of each. The first panel shows the share of advisor and student publications in all of the roughly 2000 journals indexed by EconLit. The second panel shows the share of advisor and student publications in relatively well-cited Deep Impact journals, classified by \cite{angrist2020inside}. The third panel shows the share of publications in top-6 economics journals. The data appendix lists the journals included in the second and third panels.}
	\end{flushleft}
\end{figure} 

\begin{figure}[h!]
	\centering
	\caption{Research Activity by Cohort and School}
	\includegraphics[width=\linewidth]{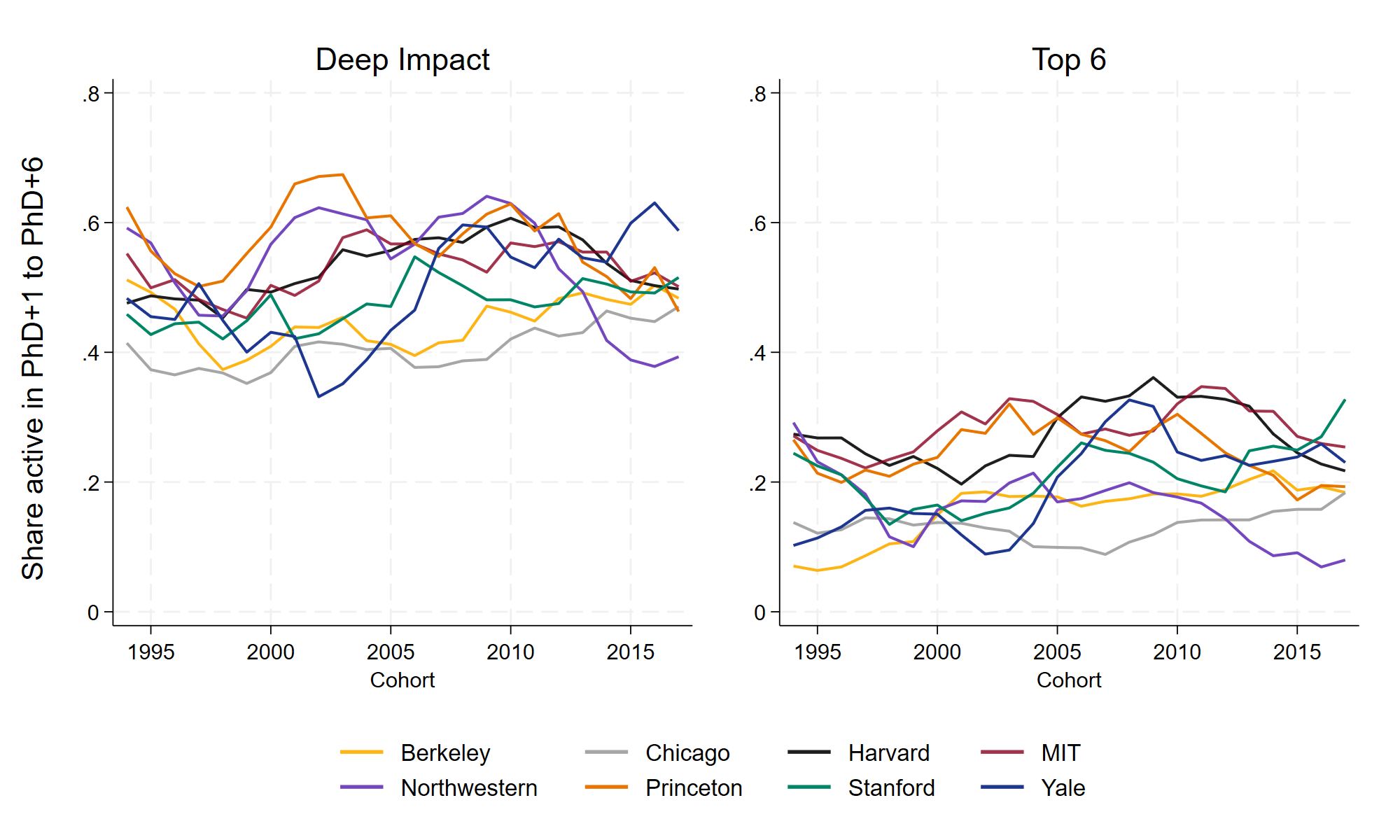}
	\label{fig: Fraction_schools_degreeyear}
        \begin{flushleft}
		\noindent\scriptsize{\emph{Notes}: This figure plots the share of graduates with at least one publication in the first six years post-PhD ($c+1$ to $c+6$) for the economics + related sample. Data for 1994-2017 graduates.}
	\end{flushleft}
\end{figure}

\begin{figure}[h!]
	\centering
	\caption{Gender Gaps in Graduate Student Research Productivity Profiles}
	\includegraphics[width=\linewidth]{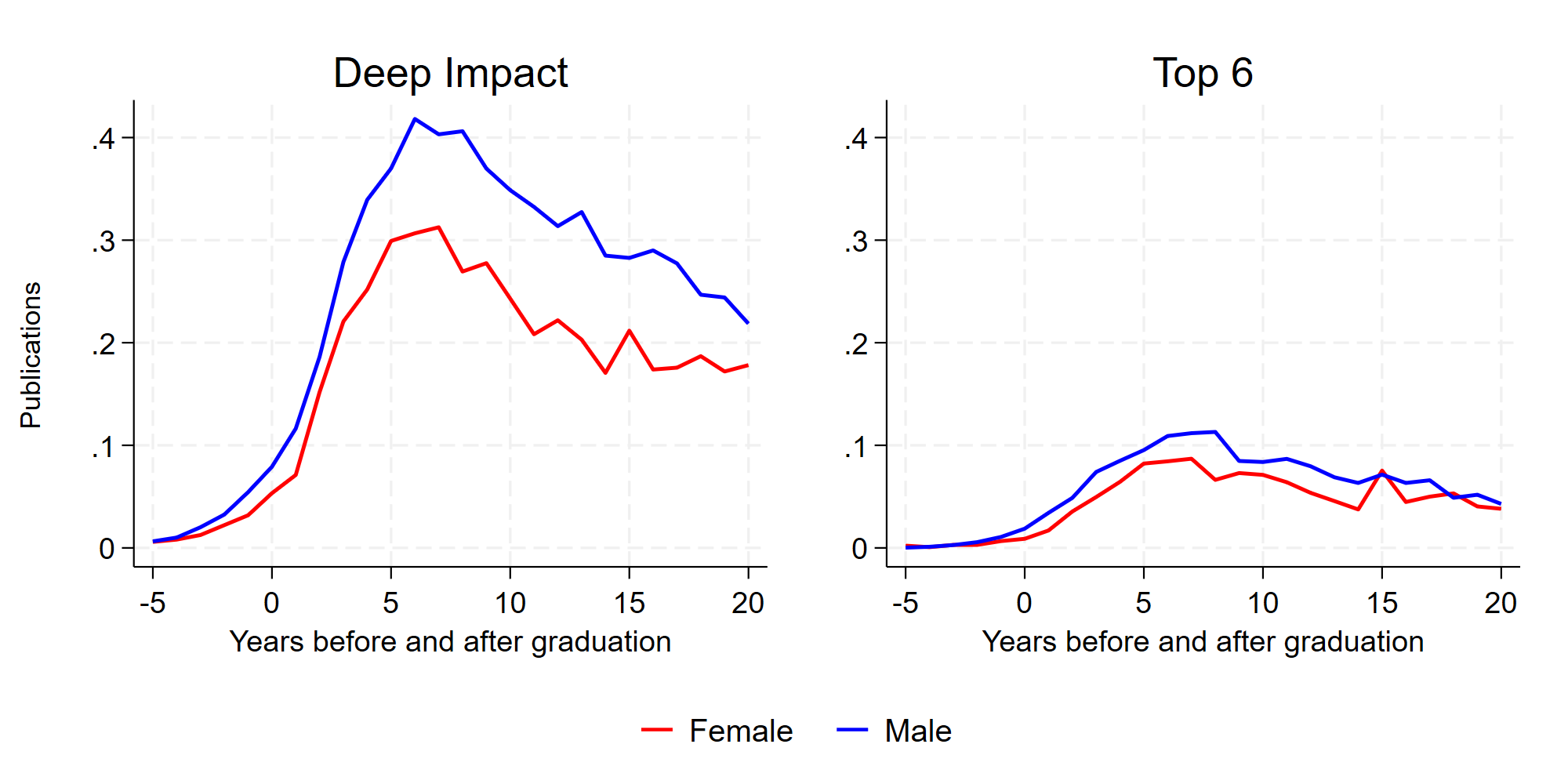}
	\label{fig: active 2 counts}
           \begin{flushleft}
		\noindent\scriptsize{\emph{Notes}: See notes to Figure \ref{fig: female shares}. This figure plots publication counts by male and female students in year $t-c$, where $c$ is graduation year.  Data are for the economics + related sample of 1994-2017 graduates. }
	\end{flushleft}
\end{figure}

\begin{figure}[h!]
	\centering
    \caption{Annual Productivity Profiles by Grad Student and Advisor Gender}	\includegraphics[width=0.8\linewidth]{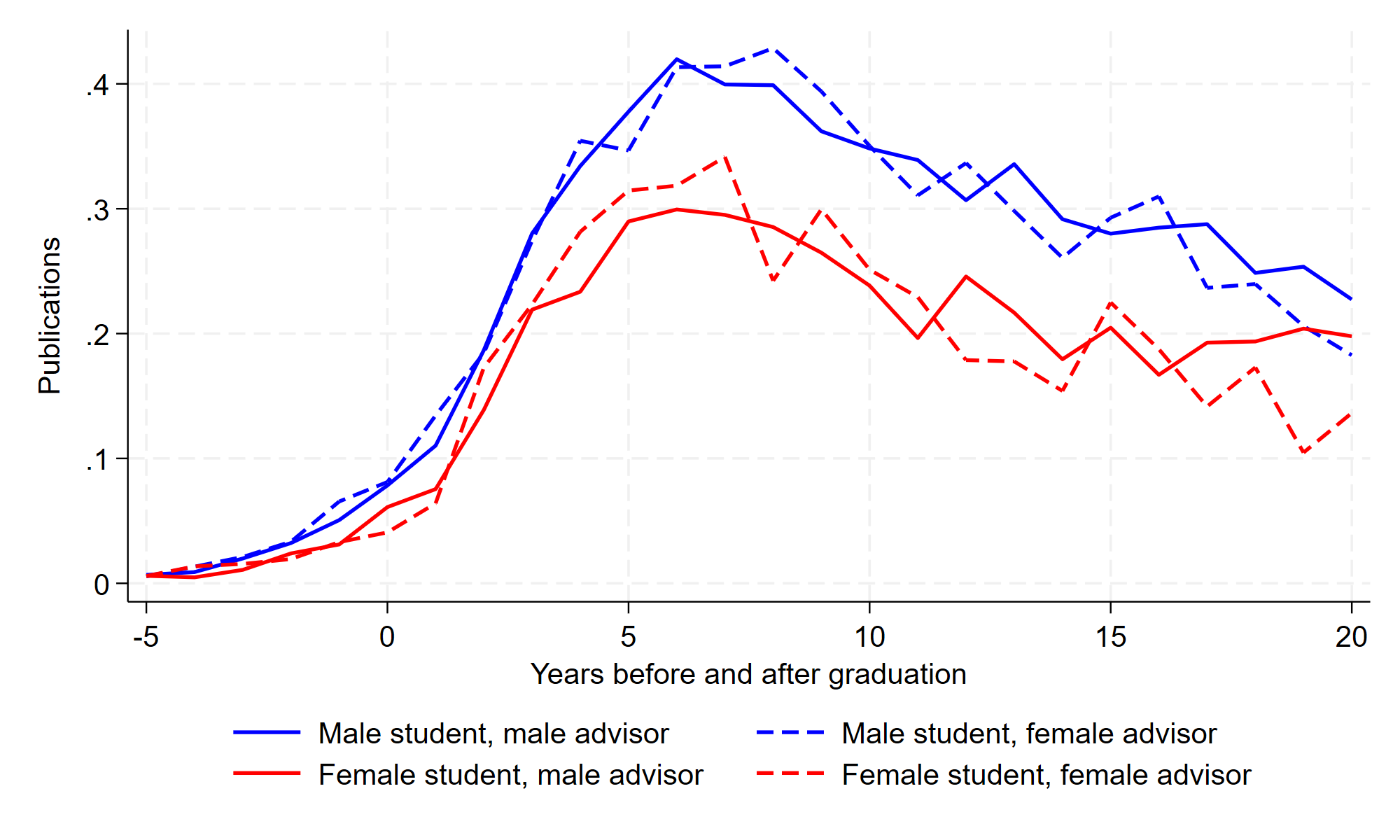}
	\label{fig: active 3 counts}
           \begin{flushleft}
		\noindent\scriptsize{\emph{Notes}:  See notes to Figure \ref{fig: female shares}. This figure plots mean publications by advisee and advisor gender in year $t-c$, where $c$ is graduation year. Graduates marked as having male advisors were advised by all-male advising teams. Graduates marked as having female advisors had at least one female advisor. Research productivity in $t$ is defined by the number of Deep Impact publications in that year. Economics + related sample of 1994-2017 graduates.}
	\end{flushleft}
\end{figure}

\begin{figure}[h!]
	\centering
	\caption{Advisor Rank for T6 and DI Student Success Measures}
	\includegraphics[width=\linewidth]{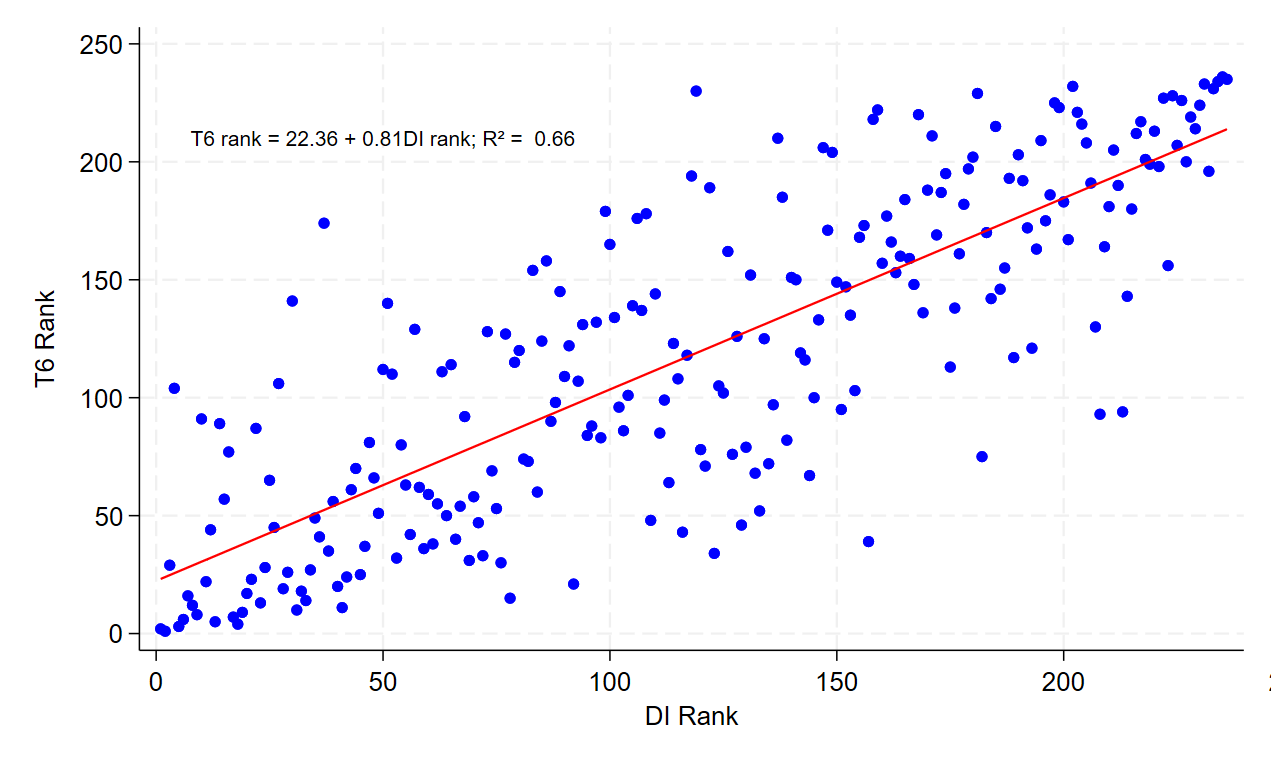}
	\label{f:rankfig}
    	\begin{flushleft}
 	\noindent\scriptsize{\emph{Notes}: This figure plots advisor rank based on T6 student success against advisor rank based on Deep Impact student success for 236 advising-loaded advisors (who advised at least 30 graduates). Both variables are computed for graduates who earned degrees at one of the eight schools in our sample in 1989-2023. Student success averages post-graduation publications in years c + 1 to c + 6 for a graduate in cohort $c$. The DI rank coefficient is highly statistically significant with a standard error of 0.034. }
	\end{flushleft}
\end{figure}

\begin{table}[h!]
    \centering
    \caption{Descriptive Statistics for the Sample Used in Tables \ref{t:studentregs}-\ref{t: Coauthoring Fixed Duper}}
    \label{t:appendixstats}
    \resizebox{0.7\columnwidth}{!}{
   \begin{tabular}{ll}
   {
\def\sym#1{\ifmmode^{#1}\else\(^{#1}\)\fi}
\begin{tabular}{l*{1}{cc}}
\hline\hline
                    &        Mean&   Std. Dev.\\
\hline
Deep Impact publications in c+1 to c+6&        1.56&        2.31\\
Top 6 publications in c+1 to c+6&        0.39&        1.00\\
Advisor Research    &        5.77&        3.79\\
Advising Load       &        8.28&        5.86\\
Past Student Success&        1.50&        0.97\\
Super Advisor Research&        0.54&        0.50\\
Super Advising Load &        0.77&        0.42\\
Super Past Student Success&        0.43&        0.50\\
Duper Advisor Research&        0.38&        0.48\\
Duper Advising Load &        0.59&        0.49\\
Duper Past Student Success&        0.20&        0.40\\
Fixed-Duper Advisor Research&        0.70&        0.46\\
Fixed-Duper Advising Load&        0.84&        0.37\\
Fixed-Duper Past Student Success&        0.77&        0.42\\
Any Publication Pre-grad&        0.10&        0.30\\
Coauthored with Advisor Pre-grad&       0.034&        0.18\\
Coauthored with Classmate Pre-grad&       0.011&        0.10\\
Number of Advisors  &        3.09&        0.73\\
Economics Dept Graduate&        0.86&        0.35\\
Unclassified Gender &        0.14&        0.34\\
No thesis PDF       &       0.069&        0.25\\
Female              &        0.27&        0.45\\
Cites Advisor in Thesis&        0.60&        0.49\\
\hline\hline
\end{tabular}
}

    \end{tabular}}
    \begin{flushleft}
        \noindent\scriptsize{\emph{Notes}: This table reports means and standard deviations for the sample of 5676 PhD students graduating  from economics and related programs between 1994-2017. The share female is calculated for 490 5students for whom gender is classified. The share citing an advisor is calculated for 5291 students with thesis PDFs.} 
    \end{flushleft}
\end{table}

\clearpage

{\footnotesize
\begin{longtable}[c]{p{0.01\textwidth} ccccccc}  
    \caption{Student Success Measures for 189 Advising-Loaded Advisors} 
\label{t:league} \\
    \toprule
    Row & Name & Avg DI  &Rank by Avg T6  & No. Advisees & Rank by No. & Affiliation \\
    \midrule
    \endfirsthead

    \toprule
   Row & Name & Avg DI  &Rank by Avg T6  & No. Advisees & Rank by No. & Affiliation \\
    \midrule
    \endhead

    \midrule
    \endfoot

    \bottomrule
    \endlastfoot

    1&Roth, Alvin E&3.82&2&60&59&Harvard\\
2&Fudenberg, Drew D&3.72&1&65&49&Harvard\\
3&Duffie, Darrell&3.32&29&41&123&Stanford\\
4&Phillips, Peter C&3.25&102&69&42&Yale\\
5&Chamberlain, Gary&3.18&3&40&126&Harvard\\
6&Poterba, James M&3.12&6&91&18&MIT\\
7&Ashenfelter, Orley C&3.05&16&44&104&Princeton\\
8&Newey, Whitney K&2.93&22&30&189&MIT\\
9&Milgrom, Paul R&2.92&12&50&84&Stanford\\
10&Andrews, Donald W&2.91&91&66&47&Yale\\
11&Imbens, Guido W&2.88&8&43&108&Harvard\\
12&Angrist, Joshua D&2.79&5&57&67&MIT\\
13&Geanakoplos, John&2.76&44&37&141&Yale\\
14&Scharfstein, David S&2.72&89&36&151&Harvard\\
15&Deaton, Angus S&2.7&77&46&95&Princeton\\
16&Dixit, Avinash K&2.67&57&33&171&Princeton\\
17&Barro, Robert J&2.65&7&54&73&Harvard\\
18&Athey, Susan C&2.62&4&42&115&Stanford\\
19&Alesina, Alberto F&2.62&9&104&14&Harvard\\
20&Hausman, Jerry A&2.61&17&57&65&MIT\\
21&Blanchard, Olivier J&2.54&23&89&22&MIT\\
22&Green, Jerry R&2.53&87&57&66&Harvard\\
23&Chay, Kenneth Y&2.49&13&35&160&Berkeley\\
24&Honore, Bo E&2.48&65&64&53&Princeton\\
25&Mankiw, Gregory&2.46&28&59&60&Harvard\\
26&Maskin, Eric S&2.44&45&82&31&Harvard\\
27&Wilson, Robert&2.43&19&44&106&Stanford\\
28&Laibson, David I&2.43&26&93&17&Harvard\\
29&Aghion, Philippe M&2.4&10&35&156&Harvard\\
30&Woodford, Michael&2.38&103&37&142&Princeton\\
31&Case, Anne C&2.37&136&54&71&Princeton\\
32&Mullainathan, Sendhil&2.37&14&52&79&Harvard\\
33&Chetty, Raj&2.33&18&51&82&Harvard\\
34&Hoxby, Caroline M&2.29&27&113&11&Stanford\\
35&Meyer, Bruce D&2.24&162&34&164&Northwestern\\
36&Watson, Mark W&2.22&55&41&122&Princeton\\
37&Rosen, Sherwin&2.22&49&41&124&Chicago\\
38&Pistaferri, Luigi&2.21&41&38&138&Stanford\\
39&Katz, Lawrence F&2.2&11&220&1&Harvard\\
40&Holmstrom, Bengt R&2.18&35&33&173&MIT\\
41&Gruber, Jonathan H&2.17&24&58&63&MIT\\
42&Kremer, Michael R&2.16&20&69&45&Harvard\\
43&Autor, David H&2.15&37&62&55&MIT\\
44&Stock, James H&2.13&50&30&183&Harvard\\
45&Grossman, Gene M&2.12&70&51&81&Princeton\\
46&Acemoglu, Daron K&2.12&25&147&5&MIT\\
47&Shleifer, Andrei&2.11&51&190&2&Harvard\\
48&Campbell, John Y&2.11&110&132&9&Harvard\\
49&Gibbons, Robert S&2.1&58&30&184&MIT\\
50&Bloom, Nicholas&2.1&80&52&77&Stanford\\
51&Powell, James&2.09&108&33&169&Berkeley\\
52&Hall, Robert E&2.07&66&42&117&Stanford\\
53&Sims, Christopher A&2.06&134&83&28&Princeton\\
54&Farber, Henry S&2.05&63&64&54&Princeton\\
55&Pakes, Ariel S&2.02&32&86&25&Harvard\\
56&Card, David E&2.01&42&159&4&Berkeley\\
57&Bernanke, Ben S&2&81&40&128&Princeton\\
58&Rogerson, William P&2&125&40&127&Northwestern\\
59&Hart, Oliver S&1.99&62&69&44&Harvard\\
60&Banerjee, Abhijit V&1.97&36&99&15&MIT\\
61&Niederle, Muriel&1.97&54&31&181&Stanford\\
62&Diamond, Peter A&1.97&59&30&187&MIT\\
63&Eckaus, Richard S&1.96&56&82&32&MIT\\
64&Greenstone, Michael B&1.96&40&82&30&MIT\\
65&Gul, Faruk R&1.96&38&53&75&Princeton\\
66&Goldin, Claudia D&1.94&47&86&23&Harvard\\
67&Horner, Johannes&1.93&113&30&188&Yale\\
68&Dekel, Eddie&1.93&109&45&100&Northwestern\\
69&Caballero, Ricardo J&1.93&31&58&62&MIT\\
70&Feldstein, Martin S&1.93&33&43&109&Harvard\\
71&Goldberg, Pinelopi K&1.92&69&38&135&Yale\\
72&Gourinchas, Pierre O&1.91&60&35&158&Berkeley\\
73&Pearce, David G&1.9&92&30&182&Yale\\
74&Whinston, Michael D&1.88&124&34&167&Northwestern\\
75&Helpman, Elhanan&1.88&53&50&86&Harvard\\
76&Werning, Ivan&1.85&15&47&93&MIT\\
77&Duflo, Esther C&1.85&30&73&38&MIT\\
78&Bernheim, B D&1.85&74&85&26&Stanford\\
79&Christiano, Lawrence J&1.85&123&65&50&Northwestern\\
80&Ely, Jeffrey C&1.84&112&45&99&Northwestern\\
81&Wolinsky, Asher&1.82&117&39&132&Northwestern\\
82&Ellison, Glenn D&1.82&73&66&48&MIT\\
83&Farhi, Emmanuel&1.81&21&32&175&Harvard\\
84&Stein, Jeremy C&1.8&148&107&13&Harvard\\
85&Xiong, Wei&1.76&151&37&140&Princeton\\
86&Shoven, John B&1.75&90&53&74&Stanford\\
87&Rabin, Matthew&1.74&120&54&72&Berkeley\\
88&Pencavel, John H&1.74&126&46&96&Stanford\\
89&Lazear, Edward P&1.73&61&30&186&Stanford\\
90&Rosenzweig, Mark R&1.72&96&36&152&Yale\\
91&Sadoulet, Elisabeth&1.72&118&71&41&Berkeley\\
92&Noll, Roger G&1.72&140&39&131&Stanford\\
93&Macurdy, Thomas E&1.71&107&42&112&Stanford\\
94&Taylor, John B&1.69&88&36&149&Stanford\\
95&Saez, Emmanuel&1.68&105&44&102&Berkeley\\
96&Altonji, Joseph G&1.68&127&90&19&Yale\\
97&Sargent, Thomas J&1.67&84&58&61&Chicago\\
98&Piazzesi, Monika&1.64&86&42&113&Stanford\\
99&Eichenbaum, Martin S&1.64&156&86&24&Northwestern\\
100&Glaeser, Edward L&1.63&48&136&7&Harvard\\
101&Rogoff, Kenneth S&1.63&83&62&56&Harvard\\
102&Gilbert, Richard J&1.63&129&35&155&Berkeley\\
103&Jones, Charles I&1.61&167&31&180&Stanford\\
104&Udry, Christopher R&1.6&135&83&27&Yale\\
105&Bresnahan, Timothy F&1.6&164&90&21&Stanford\\
106&Wolak, Frank A&1.6&94&80&33&Stanford\\
107&Tadelis, Steven&1.59&100&32&174&Berkeley\\
108&Scheinkman, Jose A&1.58&131&33&172&Chicago\\
109&Polak, Ben&1.57&166&35&159&Yale\\
110&Chiappori, Pierre A&1.56&97&34&166&Chicago\\
111&Mortensen, Dale T&1.55&138&42&111&Northwestern\\
112&Bergemann, Dirk&1.54&119&37&144&Yale\\
113&Dellavigna, Stefano&1.54&85&50&85&Berkeley\\
114&Pesendorfer, Wolfgang&1.52&64&44&105&Princeton\\
115&Delong, J Bradford&1.51&106&51&80&Berkeley\\
116&Abramitzky, Ran&1.5&43&34&165&Stanford\\
117&Rossi Hansberg, Esteban&1.5&46&42&116&Princeton\\
118&Mokyr, Joel&1.49&175&65&51&Northwestern\\
119&Finkelstein, Amy N&1.49&34&47&94&MIT\\
120&Auffhammer, Maximilian&1.49&82&39&130&Berkeley\\
121&Magruder, Jeremy R&1.49&72&37&145&Berkeley\\
122&Romer, David H&1.48&155&58&64&Berkeley\\
123&Auerbach, Alan J&1.48&115&46&97&Berkeley\\
124&Cochrane, John H&1.48&173&44&103&Chicago\\
125&Jorgenson, Dale W&1.48&122&42&118&Harvard\\
126&Antras, Pol&1.48&75&42&119&Harvard\\
127&Ferrie, Joseph P&1.47&189&38&137&Northwestern\\
128&Heckman, James J&1.47&39&75&37&Chicago\\
129&Kiyotaki, Nobuhiro&1.46&78&39&133&Princeton\\
130&Miguel, Edward A&1.46&104&111&12&Berkeley\\
131&Einav, Liran&1.46&79&61&57&Stanford\\
132&Tamer, Elie T&1.46&101&48&91&Northwestern\\
133&Melitz, Marc J&1.44&68&41&125&Harvard\\
134&Borenstein, Severin J&1.43&71&37&146&Berkeley\\
135&Kline, Patrick M&1.43&52&37&143&Berkeley\\
136&Myerson, Roger B&1.42&146&45&98&Chicago\\
137&Gorodnichenko, Yuriy&1.4&121&50&83&Berkeley\\
138&Gertler, Paul J&1.39&170&36&148&Berkeley\\
139&Becker, Gary S&1.38&145&134&8&Chicago\\
140&Dornbusch, Rudiger W&1.38&128&64&52&MIT\\
141&Obstfeld, Maurice M&1.38&95&56&68&Berkeley\\
142&Morris, Stephen E&1.37&114&68&46&Princeton\\
143&Perloff, Jeffrey M&1.36&181&72&39&Berkeley\\
144&Cutler, David M&1.35&98&161&3&Harvard\\
145&Williamson, Jeffrey G&1.34&180&61&58&Harvard\\
146&Ligon, Ethan&1.34&179&32&178&Berkeley\\
147&Abreu, Dilip&1.34&144&41&121&Princeton\\
148&Benabou, Roland J&1.34&116&44&101&Princeton\\
149&Hoynes, Hilary W&1.33&132&33&168&Berkeley\\
150&Olken, Benjamin A&1.33&67&30&185&MIT\\
151&Akerlof, George A&1.33&161&55&69&Berkeley\\
152&Nevo, Aviv&1.33&143&40&129&Northwestern\\
153&Wright, Brian D&1.3&186&69&43&Berkeley\\
154&Klenow, Peter&1.29&163&34&163&Stanford\\
155&Wolfram, Catherine&1.29&99&48&90&Berkeley\\
156&Segal, Ilya R&1.29&93&31&179&Stanford\\
157&Villas Boas, Sofia B&1.28&153&53&76&Berkeley\\
158&Mas, Alexandre&1.28&141&36&153&Princeton\\
159&De Janvry, Alain&1.27&158&75&36&Berkeley\\
160&Stavins, Robert N&1.26&165&35&154&Harvard\\
161&Anderson, Robert M&1.25&184&32&177&Berkeley\\
162&Hansen, Lars P&1.22&157&142&6&Chicago\\
163&Levitt, Steven D&1.22&147&78&34&Chicago\\
164&Freeman, Richard B&1.21&150&48&92&Harvard\\
165&Eichengreen, Barry J&1.19&152&83&29&Berkeley\\
166&Srinivasan, T N&1.19&182&37&147&Yale\\
167&Metrick, Andrew&1.19&171&32&176&Yale\\
168&Porter, Robert&1.17&169&123&10&Northwestern\\
169&Temin, Peter&1.16&142&43&107&MIT\\
170&Lee, Ronald D&1.16&160&38&139&Berkeley\\
171&Reiss, Peter C&1.15&185&33&170&Stanford\\
172&Brunnermeier, Markus K&1.15&130&48&89&Princeton\\
173&Shiller, Robert J&1.15&176&48&88&Yale\\
174&Brainard, William C&1.14&172&43&110&Yale\\
175&Alvarez, Fernando E&1.13&111&71&40&Chicago\\
176&Levin, Jonathan D&1.12&76&42&114&Stanford\\
177&Pavan, Alessandro&1.11&177&36&150&Northwestern\\
178&Hanemann, W Michael&1.1&178&50&87&Berkeley\\
179&Raphael, Steven P&1.1&133&52&78&Berkeley\\
180&Murphy, Kevin M&1.09&154&54&70&Chicago\\
181&Townsend, Robert M&1.09&159&76&35&Chicago\\
182&Krugman, Paul R&1.09&168&34&162&MIT\\
183&Karp, Larry S&1.08&188&38&136&Berkeley\\
184&Rausser, Gordon C&1.07&174&41&120&Berkeley\\
185&Rogerson, Richard D&1.06&137&35&161&Princeton\\
186&Zilberman, David&1.04&187&93&16&Berkeley\\
187&Greif, Avner&1.03&183&35&157&Stanford\\
188&Panzar, John&1.03&139&39&134&Northwestern\\
189&Lucas, Robert E&1.02&149&90&20&Chicago

\end{longtable}

\begin{flushleft}
		\noindent\scriptsize{\emph{Notes}: This table ranks prolific advisors (who advised at least 30 graduates) by the average DI publications of their advisees who earned degrees at one of the eight schools in our sample in 1989-2023, truncated to advisors for whom their students' average DI publications exceed one. Row order ranks advisors by their students' average post-graduation DI publications (i.e., in years $c+1$ to $c+6$ for a graduate in cohort $c$). The table also reports ranks by students' average T6 publications, the number of students advised, and advisor rank by number advised. The affiliation column in the table indicates the modal school from which an advisor's advisees graduated. Rankings change little when student publications exclude work coauthored with advisors.
}
\end{flushleft}
}

\newpage

\newpage

\bibliographystyle{Bibliography/aea}
{\footnotesize\bibliography{Draft/references.bib}}
 
\newpage

\end{document}